\documentclass[twocolumn,times,tighten]{aastex631}

\usepackage{amsmath}
\usepackage{physics}
\usepackage{booktabs} 
\usepackage{multirow} 
\usepackage{lineno}
\usepackage{natbib}
\usepackage{comment}
\usepackage{listings}
\usepackage{tikz}
\usetikzlibrary{arrows.meta, positioning}

\lstdefinelanguage{Julia}{
  morekeywords={using, import, function, return, end, do, begin, for, while,
    if, else, elseif, struct, mutable, const, module, export, true, false,
    nothing},
  sensitive=true,
  morecomment=[l]{\#},
  morestring=[b]{"},
}
\newcommand{\aether}{\texttt{Aether.jl}}
\newcommand{\code}[1]{\texttt{#1}}
\newcommand{\divB}{\nabla\cdot\mathbf{B}}

\newcommand{\rev}[1]{#1}

\shorttitle{Aether.jl: Methods, Design, and Validation}
\shortauthors{Ho}

\graphicspath{{./}{figures/}}

\begin{document}

\title{\aether: A High-Performance 3D MHD and Multifluid Dust Code Written in a Dynamic Language \\ with an Interactive Human-AI Development Framework}

\correspondingauthor{Ka Wai Ho}
\email{kawaiho@kitp.ucsb.edu}

\author[0000-0003-3328-6300]{Ka Wai Ho}
\affiliation{Kavli Institute for Theoretical Physics, University of California, Santa Barbara, CA 93106, USA}

\begin{abstract}
We present \aether \footnote{\url{https://github.com/MHDFlows/Aether.jl}}, a new finite-volume code for compressible hydrodynamics and magnetohydrodynamics, written in Julia and primarily designed for GPU systems.
The code solves the MHD equations with constrained transport in Cartesian, cylindrical, and spherical-polar coordinates, using standard high-order Godunov methods.
An arbitrary number of dust fluids can be coupled to the gas through stiff mutual drag.
\rev{The code was built from scratch through a human-coding agent workflow that pairs a repository-resident memory system with a self-correcting review loop.
Alongside the numerical methods, we document this framework and analyze its development record.}
Performance-critical kernel is written through \code{KernelAbstractions.jl}, and supports runs on CPUs and GPUs from multiple vendors.
\aether\ can be ran either from an interactive notebook or batch scripts, keeping prototyping, production runs, and analysis in a single language.
We verify the implementation through a series of hydrodynamic, MHD, and dust tests.
Although written in a dynamic language, \aether\ achieves comparable or even higher single-GPU throughput than C++ code on the same hardware.
\rev{In weak scaling on Frontier, \aether\ sustains 91\% parallel efficiency on 32,768 GCDs, the largest GPU run of a Julia code reported to date.}
These results show that a dynamic language now supports production astrophysical MHD simulations on exascale systems.
\aether\ and its Jupyter notebook example suite are publicly available.
\end{abstract}

\section{Introduction} \label{sec:intro}

Grid-based fluid and magnetohydrodynamics (MHD) codes form part of the core infrastructure of modern astrophysics.
Over the past three decades, community codes such as ZEUS \citep{StoneNorman1992}, FLASH \citep{Fryxell2000}, RAMSES \citep{Teyssier2002}, PLUTO \citep{Mignone2007pluto}, Athena \citep{Stone2008}, Enzo \citep{Bryan2014}, and Athena++ \citep{Stone2020} have supported research in nearly every subfield of the discipline.
Their applications range from planet formation and accretion flows to the interstellar medium, galaxy evolution, and galaxy clusters.
Nearly all of these codes are written in static compiled languages: C, C++, or Fortran.
Their designs also reflect the CPU clusters of their era, with distributed-memory parallelism through MPI and data structures tuned for per-core cache performance.

However, in the last decade, mainstream supercomputing has moved to heterogeneous architectures in which GPUs supply nearly all of the compute capability.
Current exascale systems, including Frontier, Aurora, and El Capitan, follow this design.\footnote{\url{https://top500.org}}
Efficient use of such machines requires fine-grained parallelism over millions of threads and careful control of data movement between host and device.
These requirements motivated a new generation of astrophysical codes designed for GPUs from the outset, including Cholla \citep{Schneider2015}, GAMER-2 \citep{Schive2018}, K-Athena \citep{Grete2021}, QUOKKA \citep{WibkingKrumholz2022}, Parthenon \citep{Grete2023}, IDEFIX \citep{Lesur2023}, and AthenaK \citep{Stone2026}.
Several of them achieve portability across GPU vendors through \rev{C++ abstraction libraries such as Kokkos \citep{Trott2022} and AMReX \citep{Zhang2019}}.
These codes modernize the execution model of astrophysical simulation, but they retain the language model of their predecessors.

Staying with static languages preserves a long-standing split in the research workflow, commonly called the two-language problem \citep{Bezanson2017}.
Production solvers are written in compiled languages because interpreted execution cannot meet their performance demands.
Prototyping, analysis, post-processing, and visualization instead take place in dynamic languages such as Python, MATLAB, and Julia, whose interactive sessions and rich libraries make exploratory work fast.
The split carries costs.
A scheme validated in a prototype must be rewritten in the production language before it can run at scale.
Each translation costs development time and can introduce new errors.

The same division also imposes a steep learning curve.
Programming in astronomy now starts with dynamic languages, most often Python, so a production code may be the first large compiled program a researcher encounters.
Before contributing any physics, a newcomer must absorb template-heavy C++, build systems, and GPU programming models.
Layers of performance abstraction further obscure the correspondence between the published scheme and its implementation.
For many users the production codes therefore remain black boxes, run but never read.
An ideal workflow would keep prototyping, production runs, and analysis within a single language and a single environment.

The Julia language was designed to close this gap \citep{Bezanson2017}.
Julia pairs a dynamic, mathematical syntax with just-in-time compilation to native machine code through LLVM.
Multiple dispatch specializes generic algorithms at compile time, so high-level source can reach the performance of compiled C.
The same compilation path extends to accelerators.
\code{CUDA.jl} compiles Julia functions directly into GPU kernels \citep{Besard2019}.
\code{KernelAbstractions.jl} generalizes this model: a single kernel definition targets NVIDIA, AMD, Apple, and Intel devices as well as multithreaded CPUs \citep{Churavy2021}.
\code{MPI.jl} supplies distributed-memory communication, including GPU-aware transfers \citep{Byrne2021}.
Production applications built on this stack now run at scale in neighboring fields, notably the ocean model \code{Oceananigans.jl} \citep{Ramadhan2020} and the conservation-law framework \code{Trixi.jl} \citep{Ranocha2022}.A Julia-native, GPU-first finite-volume code for compressible astrophysical MHD is therefore a promising direction that remains open.

The practice of writing scientific software is also changing at the same time.
Large language models trained on code \citep{Chen2021, Austin2021} now power interactive coding agents that work inside the terminal, such as Claude Code and Codex.\footnote{\url{https://github.com/anthropics/claude-code}; \url{https://github.com/openai/codex}}
These agents already resolve a substantial fraction of real repository issues in benchmarks such as SWE-bench \citep{Jimenez2024}.
In an agentic workflow, the developer specifies intent, reviews changes, and makes design decisions, while the agent writes much of the source.
How to conduct such a workflow across the lifetime of a large scientific code remains an open question.
Numerical solvers demand sustained correctness, physical validation, and long-term maintainability, and none of these follow automatically from code generation.
A new code built from an empty repository offers a controlled setting in which to test both the language and the workflow.

We present \aether, a finite-volume code for compressible hydrodynamics and ideal MHD that realizes this experiment.
The code is written entirely in Julia.
It was developed from scratch in an interactive workflow between the author and coding agents.
The project therefore asks two questions.
First, can a dynamic language deliver the performance and scalability expected of a production astrophysics code on modern GPU hardware?
Second, can a human-AI workflow build such a code while preserving the maintainability, readability, and verification discipline that scientific software requires?


This paper documents the design and presents evidence on both questions.
We verify the code using a standard suite of hydrodynamic and MHD test problems.
Its measured throughput is competitive with contemporary GPU codes.

The remainder of this paper is organized as follows.
Section~\ref{sec:physics} presents the numerical methods.
Section~\ref{sec:software} describes how these methods are implemented in a dynamic language and the workflow this enables.
Section~\ref{sec:workflow} describes the human-AI development framework.
Section~\ref{sec:tests} reports the validation suite.
Section~\ref{sec:performance} reports performance benchmarks.
Section~\ref{sec:applications} demonstrates two production-scale applications.
Section~\ref{sec:summary} summarizes the results and discusses future directions.

\section{Numerical Methods} \label{sec:physics}

\subsection{Newtonian Magnetohydrodynamics} \label{sec:mhd}

We solve the equations of Newtonian MHD for a compressible, conducting fluid.
In units in which the magnetic permeability is unity, they read
\begin{align}
\frac{\partial \rho}{\partial t} + \nabla\cdot\left(\rho\mathbf{v}\right) &= 0, \label{eq:mass}\\
\frac{\partial \left(\rho\mathbf{v}\right)}{\partial t} + \nabla\cdot\left(\rho\mathbf{v}\mathbf{v} - \mathbf{B}\mathbf{B} + P^{*}\mathsf{I}\right) &= \mathbf{S}_{m}, \label{eq:momentum}\\
\frac{\partial E}{\partial t} + \nabla\cdot\left[\left(E + P^{*}\right)\mathbf{v} - \mathbf{B}\left(\mathbf{B}\cdot\mathbf{v}\right)\right] &= S_{E}, \label{eq:energy}\\
\frac{\partial \mathbf{B}}{\partial t} - \nabla\times\left(\mathbf{v}\times\mathbf{B}\right) &= 0, \label{eq:induction}
\end{align}
where $\rho$ is the mass density, $\mathbf{v}$ the velocity, $\mathbf{B}$ the magnetic field, and $\mathsf{I}$ the identity tensor.
The total pressure $P^{*} = p + \mathbf{B}\cdot\mathbf{B}/2$ combines the gas pressure $p$ with the magnetic pressure.
The total energy density $E$ sums the internal, kinetic, and magnetic contributions,
\begin{equation}
E = \rho e + \tfrac{1}{2}\rho\,\mathbf{v}\cdot\mathbf{v} + \tfrac{1}{2}\mathbf{B}\cdot\mathbf{B}, \label{eq:total_energy}
\end{equation}
where $e$ is the specific internal energy.
An ideal-gas equation of state, $p = \left(\gamma - 1\right) \rho e$ with constant adiabatic index $\gamma$, closes the system.
The momentum source $\mathbf{S}_{m}$ collects the optional non-ideal terms: gravity, stochastic forcing, and the drag exerted by dust (Section~\ref{sec:dust}).
The energy source $S_{E}$ carries the work of those forces together with optically thin cooling and heating.
Both vanish for the ideal system; Section~\ref{sec:sources} describes how all source terms are applied.

In addition to the full MHD system described above, \aether\ supports two independent simplifications.
An isothermal mode fixes $p = c_{\rm s}^{2}\,\rho$ at constant sound speed $c_{\rm s}$ and omits Equation~\ref{eq:energy}.
A hydrodynamic mode sets $\mathbf{B} = 0$, which reduces Equations~\ref{eq:mass}--\ref{eq:energy} to the Euler equations; such runs allocate no magnetic storage.
The four combinations of adiabatic or isothermal closure with hydrodynamics or MHD are selected through the equation-of-state type.
For robustness near vacuum, every closure enforces a density floor during the conserved-to-primitive conversion; the ideal gas closures additionally enforce a pressure floor and can impose an optional temperature ceiling.

The equations are solved in conservative form.
Mass, momentum, and energy advance as cell averages under a finite-volume Godunov scheme: approximate Riemann solvers supply the face fluxes (Section~\ref{sec:fluxes}), and a method-of-lines integrator supplies the time discretization (Section~\ref{sec:time}).
The induction equation is discretized separately with constrained transport, which preserves $\divB = 0$ (Section~\ref{sec:ct}).

\subsection{Multifluid Dust} \label{sec:dust}

Dust is modeled as $N$ pressureless fluid species coupled to the gas by linear drag.
Each species $k$ carries its own density $\rho_k$ and velocity $\mathbf{v}_k$ and obeys
\begin{align}
\frac{\partial \rho_k}{\partial t} + \nabla\cdot\left(\rho_k \mathbf{v}_k\right) &= 0, \label{eq:dust_mass}\\
\frac{\partial \left(\rho_k \mathbf{v}_k\right)}{\partial t} + \nabla\cdot\left(\rho_k \mathbf{v}_k \mathbf{v}_k\right) &= \frac{\rho_k \left(\mathbf{v} - \mathbf{v}_k\right)}{t_{{\rm s},k}}, \label{eq:dust_momentum}
\end{align}
where $t_{{\rm s},k}$ is the stopping time of species $k$.
The drag term accelerates each species toward the gas velocity on that timescale.
The gas momentum equation gains the opposite reaction, $-\sum_k \rho_k \left(\mathbf{v} - \mathbf{v}_k\right)/t_{{\rm s},k}$, so the mixture conserves total momentum.
The gas energy equation gains the matching term, chosen so that the energy dissipated by the drift between phases appears as gas heat.
The stopping time is specified per species, either as a constant or as a function of position, gas density, dust density, and sound speed.
This covers Epstein, constant, and constant-coefficient drag laws \citep{Epstein1924, LehmannWardle2018}.

Each species advances through the same finite-volume machinery as the gas.
The reconstruction schemes are shared, and fluxes come from a pressureless Riemann solver: by default the diffusive Rusanov flux of \citet[Appendix E]{Krapp2024}, with an HLL variant using signed wave-speed bounds as an alternative.
The diffusive default passes robustly through the $\delta$ shocks that converging pressureless flow forms.
Cells whose dust density falls below a floor are reset to the floor with zero velocity, which keeps evacuated regions from binding the timestep.

The drag coupling is stiff whenever $t_{{\rm s},k} \ll \Delta t$, as holds for small grains.
\aether\ therefore integrates drag implicitly inside the time integrator, following \citet{Krapp2024}.
Because every species couples to the gas but not to the other species, the per-cell drag matrix has arrowhead structure and inverts in closed form at cost linear in $N$.
The implicit stage is solved exactly, conserves the total momentum of the system, and remains stable for arbitrarily small stopping times.
Section~\ref{sec:time} describes the surrounding implicit-explicit integrator.
Comparable multifluid dust schemes are implemented in FARGO3D \citep{BenitezLlambay2019} and Athena++ \citep{HuangBai2022}.

\subsection{Mesh, Coordinates, and Boundary Conditions} \label{sec:mesh}

\aether\ adopts the MeshBlock domain decomposition introduced by Athena++ \citep{Stone2020}.
The computational domain, in one, two, or three dimensions, is divided into logically identical blocks of fixed size.
Each block is padded with the maximum ghost width required by its reconstruction and Riemann solver: 1 cell for donor-cell reconstruction, 2 for PLM, and 3 for the PPM-family and WENO schemes; multidimensional LHLLC/LHLLD independently requires at least 2.
Blocks are ordered along a Morton space-filling curve and distributed over MPI ranks in contiguous slices of that ordering.
The grid is currently uniform and single-level.
The block structure is naturally open for a possible future extension of mesh refinement.

The array layout is chosen for GPU parallelism.
All cell-centered fields of a rank live in five-dimensional arrays indexed $(i, j, k, v, m)$, where $(i, j, k)$ are the cell indices, $v$ the variable, and $m$ the block.
\rev{This layout keeps all blocks on a rank contiguous in memory, analogous to how AthenaK \citep{Stone2026} modernizes the separate per-block allocations of Athena++.}
Julia stores arrays in column-major order, so the cell index $i$ varies fastest and consecutive GPU threads touch consecutive memory.
The block index is a dimension of the kernel launch itself: one launch advances every block a rank owns.
Device utilization is therefore set by the total cell count of the rank rather than by the size of any single block, which keeps small blocks viable as units of decomposition.

Three coordinate systems are supported, Cartesian $(x, y, z)$, cylindrical $(R, \phi, z)$, and spherical-polar $(r, \theta, \phi)$, each with uniform spacing in every coordinate.
Geometry enters the solver only through exact face areas, cell volumes, edge lengths, and volume-weighted cell centroids.
These factors are evaluated on demand by coordinate-specialized functions instead of being stored in arrays.
The momentum source terms required in curvilinear coordinates are described in Section~\ref{sec:sources}.

Ghost cells are filled by one exchange that covers all face, edge, and corner neighbors, up to 26 per block in three dimensions, as constrained transport requires.
Neighbors on the same rank copy block to block; remote strips are grouped by peer rank, yielding at most one send and one receive per peer in each exchange.
Physical boundary conditions are imposed after the exchange.
Built-in choices are periodic, outflow, reflecting, and a polar condition for the spherical axis, together with user-defined conditions.
The polar condition fills axis ghost zones from the cells half a turn away in azimuth, with the sign flips and the face-field and axis electric-field treatment that MHD requires.

\subsection{Reconstruction and Riemann Solvers} \label{sec:fluxes}

Within each block the conservative equations are discretized in integral form.
For a cell of volume $V$ the cell-averaged state $\bar{U}$ evolves as
\begin{equation}
\frac{d \bar{U}}{d t} = -\frac{1}{V} \sum_{f} A_{f}\, \hat{F}_{f} + \bar{S}, \label{eq:fv_update}
\end{equation}
where the sum runs over the faces of the cell, $A_{f}$ is the face area, $\hat{F}_{f}$ is the outward-directed numerical flux through that face, and $\bar{S}$ collects source terms (Section~\ref{sec:sources}).
Because $A_{f}$ and $V$ are the exact geometric quantities, Equation~\ref{eq:fv_update} applies unchanged in all three coordinate systems.
Fluxes are evaluated from the current stage state, and temporal accuracy comes from the Runge-Kutta integrator (Section~\ref{sec:time}).

Numerical fluxes are computed direction by direction.
Along each direction, cell-centered primitive variables are reconstructed \rev{componentwise} to the left and right sides of every face\rev{, alongside the cell-centered magnetic field in MHD}.
Six base schemes are available: piecewise-constant donor cell; a piecewise-linear method (PLM) with the monotonized harmonic-mean limiter of \citet{vanLeer1974}; the piecewise-parabolic method \citep[PPM;][]{ColellaWoodward1984}; PPM5, which limits a 5 point $4^{th}$-order upwind-biased interface value with the unmodified PPM limiter; $5^{th}$-order WENO-Z \citep{Borges2008}; and the adaptive-order WENO-AO(5,3) of \citet{Balsara2016}.
WENO-AO hybridizes the quartic over the full 5 cell stencil with 3 quadratic sub-stencils.
Smooth flow recovers the quartic, including at the extrema where PPM clips; discontinuities reduce the scheme to a $3^{rd}$ order WENO of the sub-stencils.
\rev{In both WENO schemes, the nonlinear weights use a scale-invariant regularization that remains well behaved in single precision \citep[cf.][]{Don2022}.}\footnote{\rev{The unnormalized weights are $\alpha_k = d_k [1 + (\tau / (\beta_k + \varepsilon))^2]$, with sub-stencil smoothness indicators $\beta_k$ and global indicator $\tau$. We set $\varepsilon = \epsilon_{\rm mach}\tau + \eta_{\min}$, where $\epsilon_{\rm mach}$ and $\eta_{\min}$ are the machine epsilon and minimum normal value of the working precision.}}
For the energy-evolving closures, both WENO schemes also come in variants augmented with the positivity-preserving rescaling of \citet{ZhangShu2010}.
\rev{These positivity-limited variants reconstruct $\mathbf{B}$ with PPM instead, because overshoots in the magnetic field can seed oscillatory induction fluxes.}

The reconstructed states are rotated into the frame of each face and passed to an approximate Riemann solver.
For hydrodynamics the options are Rusanov, HLLE \citep{HartenLaxvanLeer1983, Einfeldt1988}, and HLLC \citep{Batten1997, Toro2009}.
For MHD the options are Rusanov, HLLE, and HLLD \citep{MiyoshiKusano2005}.
HLLC and HLLD also come in the low-dissipation variants LHLLC and LHLLD of \citet{Minoshima2021}, which rescale the pressure dissipation of the contact wave with the local Mach number and suppress the numerical shock instability at grid-aligned shocks.
Section~\ref{sec:tgv} quantifies the low-Mach behavior.
Isothermal counterparts exist for every solver except HLLC and LHLLC; the isothermal HLLD follows \citet{Mignone2007iso}.

\subsection{Constrained Transport} \label{sec:ct}

The primary representation of the magnetic field is the set of area-averaged normal components on cell faces.
Cell-centered values, needed for reconstruction and for the total energy, are averages of the two bounding faces.
The face field advances by constrained transport \citep{EvansHawley1988}, which discretizes the induction equation in integral form.
Each face average $\bar{B}_{f}$ obeys
\begin{equation}
\frac{d \bar{B}_{f}}{d t} = -\frac{1}{A_{f}} \oint_{\partial f} \mathbf{E}\cdot d\boldsymbol{\ell}, \label{eq:ct}
\end{equation}
where $\mathbf{E} = -\mathbf{v}\times\mathbf{B}$ is the electric field and the circulation is a signed sum of edge-centered electric fields weighted by exact edge lengths.
Every edge is shared by two faces of any given cell, and their circulations traverse it in opposite senses.
The discrete magnetic flux leaving each cell is therefore constant in time.
As a result $\divB = 0$ holds in every supported coordinate system \citep[cf.][]{Toth2000}, and the induction equation needs no geometric source terms.
The face field carries its own register pair and joins the stage combinations of Section~\ref{sec:time}.

The edge electric fields derive from the Riemann fluxes.
In the face-frame Riemann problem the normal magnetic field takes the stored face value on both sides, and the tangential components of the resulting induction flux are face-averaged electric fields.
The four face values around each edge are combined with the upwind averaging of \citet{GardinerStone2005, GardinerStone2008}, which selects gradient corrections by the direction of mass flux across the intervening faces.
On meshes that include the spherical axis, the axis-aligned electric field is replaced by its azimuthal mean.

\subsection{Time Integration} \label{sec:time}

Explicit time integration uses Runge-Kutta methods in the two-register, low-storage form of \citet{Ketcheson2010}.
Only two copies of the conserved state exist, whatever the stage count.
Writing the registers as $u_{1}$ and $u_{2}$, stage $l$ performs
\begin{align}
u_{2} &\leftarrow u_{2} + \delta_{l}\, u_{1}, \nonumber\\
u_{1} &\leftarrow \gamma_{0,l}\, u_{1} + \gamma_{1,l}\, u_{2} + \beta_{l}\, \Delta t\, L\!\left(u_{1}\right), \label{eq:two_register}
\end{align}
where $L$ is the discrete right-hand side of Equation~\ref{eq:fv_update} and $\left(\gamma_{0,l}, \gamma_{1,l}, \beta_{l}, \delta_{l}\right)$ are the coefficients of the chosen method.
Three \rev{purely explicit} integrators are provided.
\code{RK2} and \code{RK3} are the optimal two-stage second-order and three-stage third-order strong-stability-preserving schemes \citep{Gottlieb2009}; both have unit Courant limit.
\code{RK4} is the four-stage fourth-order low-storage method of \citet{Ketcheson2010}, with Courant limit 1.39.

\rev{When dust couples to the gas through drag, the entire gas-dust system is advanced instead by the second-order implicit-explicit scheme \code{IMEX2P}.
This is the two-solve IMEX pair of \citet{AscherRuuthSpiteri1997} on which \citet{Krapp2024} build their multifluid drag solver.
\code{IMEX2P} partitions the equations by term rather than by fluid: hyperbolic fluxes and non-stiff sources form its explicit part, and drag alone forms its implicit part.
The explicit part is a two-stage scheme in the same two-register form as Equation~\ref{eq:two_register}; the implicit part is diagonally implicit, stiffly accurate, and L-stable.
Each stage first applies Equation~\ref{eq:two_register} to the gas and every dust species from the same frozen stage state.
One closed-form drag solve (Section~\ref{sec:dust}) then updates the gas and dust momenta and gas energy together with implicit weight $\lambda\,\Delta t$, where $\lambda = 1 + 1/\sqrt{2}$.}
This choice keeps every explicit combination weight nonnegative; the admissible Courant fraction of the explicit part becomes $1/\lambda \approx 0.586$.
Drag is woven into the stages rather than operator split because splitting fails to reach the correct drift equilibrium when external forces act \citep[Appendix C of][]{Krapp2024}.

The timestep follows from the fastest signal speed crossing each cell.
After every cycle the code evaluates
\begin{equation}
\Delta t = C\, C_{\max} \min_{\rm cells}\, \min_{d}\, \frac{\Delta x_{d}}{\left|v_{d}\right| + c_{d}}, \label{eq:cfl}
\end{equation}
where $\Delta x_{d}$ is the physical cell width along direction $d$ (an arc length in curvilinear coordinates) and $v_{d}$ is the flow velocity along $d$.
The signal speed $c_{d}$ is the sound speed in hydrodynamics, the directional fast magnetosonic speed in MHD, and zero for pressureless dust.
The factor $C_{\max}$ is the stability limit of the chosen integrator, and the Courant factor $C \in (0, 1]$ sets the fraction of that limit to run at, with default $C = 0.3$.
Because the minimum is taken per direction rather than summed over directions, multidimensional runs typically use $C \approx 1/n_{\rm dim}$.
Source terms may declare local timescales that enter the same reduction.
\rev{The timestep is independent of the dust stopping times $t_{{\rm s},k}$.
Species with $t_{{\rm s},k} \gtrsim \Delta t$ are resolved at second order by \code{IMEX2P}, and stiff species ($t_{{\rm s},k} \ll \Delta t$) relax onto their terminal-drift equilibrium without resolving the initial transient (Section~\ref{sec:dustdamp}).}

\subsection{Source Terms} \label{sec:sources}
\aether\ designed a flexible framework for cell-centered source terms. Cell-centered source terms beyond the ideal fluxes enter through three common integration slots: unsplit terms act within the Runge-Kutta stages, split terms after the stage loop, and driving terms before it.

Pointwise sources are functions of position, time, and the local primitive state of every fluid; each returns rates of change for the conserved variables it affects.
Built-in pointwise packages and user-defined terms take this form, while components that need face fluxes, global reductions, or nonlinear solves implement the same slots directly.
Each slot serves a different class of process, ansd the built-in packages illustrate the match.
A source may also declare a local timescale, which then enters the timestep reduction of Equation~\ref{eq:cfl}.

Unsplit sources, the default, are added at every Runge-Kutta stage with the same stage weight $\beta_{l}\,\Delta t$ as the flux divergence, so time-independent, state-dependent sources participate in the full-order Runge-Kutta update; explicitly time-dependent rates are currently evaluated at the cycle-start time because stage abscissae are not tracked.
All unsplit sources fuse into one kernel, which gathers the local state once and accumulates every rate before storing.
This mode fits forces that are smooth on the step and should track the flow through every stage.


The curvature terms of cylindrical and spherical-polar coordinates are applied unsplit at every stage.
\aether\ builds them from the total stress tensor
\begin{equation}
T_{ij} = \rho\, v_{i} v_{j} + P^{*} \delta_{ij} - B_{i} B_{j}, \label{eq:stress}
\end{equation}
whose divergence in curvilinear coordinates produces the momentum sources, for example $\rho v_{\phi}^{2}/R$ in cylindrical geometry; hydrodynamics is the $\mathbf{B} = 0$ case of the same expression.
The discrete factors of these sources are the same face areas and cell volumes as the flux divergence, so a uniform medium at rest stays at rest to machine precision.
The terms attach automatically on curvilinear meshes and apply in pressureless form to each dust species.

Split sources apply once per step, after the stage loop, as a single first-order operator-split update over the full $\Delta t$.
The mode gives up temporal order in exchange for stability and cost, and an optional interval super-cycles a term over several steps with the accumulated weight.

Driving sources apply once per step, before the integrator begins, so the injected field stays frozen across the stages of that step.
This slot serves stochastic forcing, for which the built-in turbulence driver is the model case.
It forces the large-scale Fourier modes of a periodic box with an Ornstein-Uhlenbeck process of prescribed correlation time \citep{EswaranPope1988, Schmidt2009}.
A Helmholtz projection of the mode amplitudes sets the solenoidal fraction of the forcing.
Each kick subtracts the mean acceleration, so no net momentum enters, and rescales the amplitude so the injected energy matches the prescribed power exactly.
Kicks may be applied every step or accumulated into impulses at a fixed driving interval.

In addition, magnetic sources can be entered as user-defined electric fields added at cell edges inside the constrained-transport update, which preserves $\divB = 0$ by construction.
For Dust-Gas coupling, as described in Section~\ref{sec:time}, stiff drag bypasses the source handler altogether: it is integrated by the implicit stages of \code{IMEX2P}.

\subsection{Dual-Energy Formalism} \label{sec:dualenergy}

Supersonic and strongly magnetized flows store nearly all of their energy in kinetic and magnetic form, which makes the total energy an ill-conditioned source for the thermal pressure.
Recovering the internal energy from Equation~\ref{eq:total_energy} then subtracts nearly equal numbers, so the truncation error of the scheme can become comparable to the internal energy itself.
Since pressure forces are negligible in exactly these regions, the error leaves the dynamics unaffected.
Nevertheless, whenever the temperature itself is required (e.g., for chemistry or radiative cooling), a remedy is needed.
For these regimes \aether\ implements the dual-energy formalism of \citet{Bryan1995}, in the entropy form adopted by GAMER-2 \citep{Schive2018}.

The formalism evolves a modified entropy density alongside the total energy.
In smooth adiabatic flow it obeys
\begin{equation}
\frac{\partial S}{\partial t} + \nabla\cdot\left(S\,\mathbf{v}\right) = 0, \qquad S = \frac{p}{\rho^{\gamma-1}}, \label{eq:dual_entropy}
\end{equation}
so the pressure follows from $S$ and the density alone, with no subtraction.
At shocks the physical entropy jumps while Equation~\ref{eq:dual_entropy} does not, so the advected value is trusted only where dissipation is absent.
In the discrete update, $S$ is carried as a passively advected scalar.
Every Riemann solver leaves its dynamical flux unchanged and adds an entropy flux equal to the mass flux times the entropy per unit mass, $S/\rho$, of its upwind side \citep{Larrouturou1991}.

The total energy and the advected entropy now provide two estimates of the pressure, and the conserved-to-primitive conversion chooses between them cell by cell.
The selection uses the ratio of the recovered internal energy to the kinetic plus magnetic energy (kinetic alone in hydrodynamics), which controls the relative error of the subtraction.
Below a threshold $\alpha_{\rm switch}$ ($10^{-3}$ by default) the subtraction is no longer trusted: the internal energy comes from the entropy, $\rho e = S\rho^{\gamma-1}/\left(\gamma-1\right)$, and the stored total energy is rewritten to match.
\rev{This step sacrifices exact energy conservation in the switched cells.}
Above a second threshold $\alpha_{\rm sync}$ ($10^{-1}$ by default) the subtraction is accurate, so the conservative value stands and $S$ is resynchronized from it.
This resynchronization alone carries shock heating into the entropy, so no separate shock detector is needed.
Between the thresholds the conservative energy still sets the pressure but $S$ is left untouched.
The entropy is thus written only where the subtraction is accurate and read only where it has failed, so its error never returns to the pressure.

\section{Implementation in a Dynamic Language} \label{sec:software}

In this section we describe how features native to a dynamic language shape the implementation of \aether\ as a production solver.
Rather than adopting an architecture inherited from established static-language codes, we organize each part of the program around a focused Julia abstraction.
Our goal is to retain the performance required for production calculations while keeping the implementation concise and intuitive.
The same principle governs both the computational core---numerical specialization, portable CPU/GPU kernels, and distributed communication---and the user-facing layers for problem setup and data output.
Sections~\ref{sec:dispatch} through \ref{sec:mpi} present the computational components, while Sections~\ref{sec:ui} and \ref{sec:io} describe the user-facing interface and output system.

\subsection{Multiple Dispatch} \label{sec:dispatch}

Scientific simulation codes must support many combinations of physical models, numerical schemes, boundary treatments, and hardware backends.
In conventional Fortran or C implementations, these choices are often selected through \code{if/else} branches or preprocessor flags such as \code{\#ifdef}.
As features accumulate, branches spread into inner loops, build configurations multiply, and the active code path becomes difficult to trace.
\rev{Modern C++ codes such as Parthenon \citep{Grete2023}, QUOKKA \citep{WibkingKrumholz2022}, and AthenaK \citep{Stone2026} eliminate these inner-loop branches through templates and static function overloading.}
\rev{In a conventional ahead-of-time C++ build, runtime options select among kernel specializations compiled into the executable.}
\rev{\aether\ reaches the same branch-free specialization directly from runtime values through what we call implicit propagation.}
The user specifies each choice once, for example by passing \code{RK3()} or \code{HLLD()} to a constructor, and Julia specializes every routine that depends on it.
The remainder of this subsection explains how multiple dispatch and compilation make this propagation possible.

Julia's multiple dispatch provides the mechanism behind this implicit propagation.
A Julia function can have several methods, and the method used for a particular call is selected from the \rev{runtime} types of all its arguments \citep{Bezanson2017}.
\rev{This form of dispatch originated in the Common Lisp Object System \citep{Bobrow1988}.}
\rev{C++ function overloading instead resolves calls from static compile-time types, and interactive C++ interpreters such as Cling in ROOT keep the same static dispatch rules \citep{Brun1997, Vasilev2012}.}
\rev{In Julia, once the runtime argument types at a call site are known, the just-in-time compiler infers types across the call tree and compiles a specialized method \citep{Bezanson2018}.}
Each interchangeable component in \aether, such as a reconstruction scheme, Riemann solver, equation of state, or time integrator, is represented by a distinct Julia type.
Values such as \code{WENOZ()}, \code{HLLD()}, and \code{RK3()} therefore identify both the user's choice and the methods that implement it.
Replacing a component consequently changes every routine that dispatches on its type without changing the corresponding call sites.
Construction then carries these type-based choices through the full solver.

The \code{Simulation} constructor does this by encoding the selected components as type parameters, so each combination of equation of state, reconstruction scheme, Riemann solver, and time integrator gives the simulation a distinct concrete type.
When \code{run!} is called for the first time, Julia compiles the methods required by that type into machine code.
The resulting kernels contain only the selected algorithms, so their inner loops need no runtime branches to determine which configuration is active.
The type information also exposes unsupported combinations before the physical state is advanced.
For example, pairing HLLC with an isothermal equation of state raises a missing-method error because no HLLC method is defined for that closure (Section~\ref{sec:fluxes}).

This specialization is not limited to numerical schemes.
Geometry-dependent operations dispatch on the coordinate type of the \code{Mesh}, allowing the same update kernels to work in every coordinate system introduced in Section~\ref{sec:mesh}.
On a spherical-polar mesh, for example, the cell volume used in Equation~\ref{eq:fv_update} is $\tfrac{1}{3}\left(r_{+}^{3} - r_{-}^{3}\right)\left(\cos\theta_{-} - \cos\theta_{+}\right)\Delta\phi$.
The azimuthal cell width used in Equation~\ref{eq:cfl} is $r\sin\theta\,\Delta\phi$.
On a Cartesian mesh, the same methods return constant grid spacings, which the compiler folds into the compiled kernels.
Selecting the coordinate system at mesh construction therefore supplies the correct geometric factors throughout the solver without modifying the update kernels or adding geometry-dependent branches.
Each compiled kernel contains only the geometry operations selected by the mesh type, so support for the other coordinate systems adds no cost to that run.

The multiple-dispatch design also makes the cost of extension predictable.
Adding a Riemann solver requires only a new type and one \rev{\code{riemann\_flux}} method.
No existing kernel, driver, or configuration file needs to change, and no \rev{branch or template instantiation table} is added to select the solver.
\rev{Julia's method tables remain open at runtime, so a user can define that type and method in a script and pass it directly to \code{Simulation} without rebuilding \aether.}
Each scheme therefore remains localized in the source as one type and its methods, keeping the implementation close to the published algorithm.

\subsection{XPU Kernels} \label{sec:xpu}
The pressure behind this layer comes from the hardware, where nearly all compute capability now sits in GPUs and each vendor exposes it through a different programming model: CUDA for NVIDIA, HIP for AMD, SYCL for Intel.
Writing kernels in one of these models ties the code to one vendor, and maintaining several kernel sets multiplies the work.
\rev{C++ codes address this through portability libraries such as Kokkos \citep{Trott2022}, RAJA \citep{Beckingsale2019}, and AMReX \citep{Zhang2019}, which compile one kernel source across multiple hardware vendors.}
\rev{K-Athena, Parthenon, IDEFIX, and AthenaK build on Kokkos \citep{Grete2021, Grete2023, Lesur2023, Stone2026}, and QUOKKA builds on AMReX \citep{WibkingKrumholz2022}.}
\rev{In conventional CUDA and HIP builds, Kokkos or RAJA kernels are ordinary C++ lambdas or functors compiled ahead of time.
Helper functions generally require device annotations, such as \code{KOKKOS\_INLINE\_FUNCTION}.
External libraries must likewise supply device-callable implementations.}

\rev{Julia instead runs the device compiler inside the language runtime.}
\code{CUDA.jl} established that the language compiler can emit GPU kernels directly from Julia functions \citep{Besard2019}.
\code{KernelAbstractions.jl} generalizes that capability into a portability layer \citep{Churavy2021, Godoy2023}: a kernel written once against its interface compiles to NVIDIA CUDA, AMD ROCm, Intel oneAPI, and Apple Metal devices, as well as to multithreaded CPU code.
\rev{Only the entry function carries the \code{@kernel} macro; every function it calls can be an ordinary, unannotated Julia function.
The device compiler specializes the full call tree at first launch, provided the called operations are supported on the target device.
This allows users to extend the solver's GPU computation with functions defined in scripts or imported from independent Julia packages written without GPU annotations.
Such composition supports the integration of additional physical models and numerical methods discussed in Sections~\ref{sec:ui} and \ref{sec:summary}.}

Every performance-critical loop in \aether\ is written against this layer, so one code path serves every architecture.
The user selects an execution backend when constructing the mesh, and that choice becomes part of the mesh's concrete type.
Passing \code{CPU()} allocates the state arrays in host memory and runs the kernels with multithreaded CPU execution.
Passing \code{GPU(CUDABackend())} instead allocates the arrays in NVIDIA GPU memory and compiles the same kernel functions for CUDA; corresponding backend wrappers target devices from the other vendors.
Array allocation and kernel launch dispatch on this backend type, so the numerical algorithms and their call sites contain no architecture-specific branches.

Nearly every computational kernel traverses the mesh data through one of two helper functions.
\rev{\code{for\_each!} executes a kernel body in parallel over a four-dimensional index range \code{(ri, rj, rk, 1:nblocks)} of cells, faces, or edges, passing the index tuple \code{(i, j, k, m)} to the closure.
The closure captures the ghost-padded arrays from its enclosing scope, so each thread can read any stencil neighborhood around \code{(i, j, k)} while writing only to its own output entry.}
\code{map\_reduce} performs a similar traversal while combining the resulting values, for example to find the timestep minimum in Equation~\ref{eq:cfl} or compute volume integrals for history output (Section~\ref{sec:io}).
These two helpers centralize launch configuration, index calculations, and bounds handling, leaving each computational kernel to specify only the operation performed on the data.

\begin{figure}[t]
\begin{lstlisting}
for_each!(architecture, ri, rj, rk,
          1:nblocks; workgroup) do i, j, k, m
    wl, wr = face_states(scheme, dir,
                         W, i, j, k, m, nv)
    f = riemann_flux(solver, eos,
                     to_face_frame(dir, wl),
                     to_face_frame(dir, wr))
    store_state!(F, from_face_frame(dir, f),
                 i, j, k, m)
end
\end{lstlisting}
\caption{The actual code of the flux kernel of the Godunov update in the \code{Hydro} module.
\code{for\_each!} maps the body over the \rev{face} ranges \code{ri}, \code{rj}, \code{rk} and all blocks of the rank, launching on the selected \code{architecture} with the given \code{workgroup} size (Section~\ref{sec:mesh}).
The body reconstructs the two face states from the primitive array \code{W}, solves the Riemann problem in the face frame, and stores the flux in \code{F}.
Because \code{scheme}, \code{solver}, and \code{eos} are the compile-time types of Section~\ref{sec:dispatch}, this one body serves every direction, reconstruction scheme, Riemann solver, equation of state, and backend.}
\label{fig:flux_kernel}
\end{figure}

The traversal helpers and multiple dispatch together collapse the entire family of Godunov flux kernels into the roughly ten lines shown in Figure~\ref{fig:flux_kernel}.
\rev{Here \code{(ri, rj, rk)} spans the faces normal to \code{dir}, where index \code{(i, j, k)} addresses the face between cell \code{(i, j, k)} and its predecessor along \code{dir}.
Dispatched on \code{scheme} and \code{dir}, \code{face\_states} reads the stencil cells along that axis from \code{W} (for example, \code{i-3} to \code{i+2} in $x_{1}$, or the matching \code{j} or \code{k} offsets).
It returns \code{wl} and \code{wr} as register tuples, which pass directly to \code{riemann\_flux} and \code{store\_state!} without intermediate global arrays.
Encoding \code{dir}, \code{scheme}, \code{solver}, and \code{eos} as concrete types (Section~\ref{sec:dispatch}) lets the compiler specialize this general body into code as concrete as a hand-written kernel.}

\rev{Backend portability also preserves the interactive workflow.
Users can develop and debug kernels on the CPU, then run them unchanged on a GPU within the same Julia session.
Switching backends is a change of one argument rather than a rebuild.}

\subsection{Parallelism} \label{sec:mpi}

Message passing remains the portable substrate for distributed memory on production clusters, and \aether\ adopts it directly.
\code{MPI.jl} wraps the system MPI library in a native Julia interface with negligible overhead \citep{Byrne2021}.
Each MPI rank owns a contiguous slice of the Morton-ordered block list (Section~\ref{sec:mesh}) and, on GPU systems, drives one device.

Communication is confined to three operations.
Ghost-zone exchange fills block boundaries with one message per pair of off-rank neighbors, as described in Section~\ref{sec:mesh}.
The timestep of Equation~\ref{eq:cfl} completes with one global reduction per cycle.
The history diagnostics of Section~\ref{sec:io} reduce their sums and extrema, conserved totals such as mass and the maximum $|\divB|$, across ranks before rank zero writes.
Every other operation in the solver is local to a rank.
Monitoring conservation on the fly therefore costs one scalar reduction per output.

On GPU systems the exchanged data lives in device memory.
With a GPU-aware MPI library, \code{MPI.jl} hands device buffers to the library directly, so messages move between GPUs without a staging copy through the host \citep{Byrne2021}.

In addition, we noted that the small-scale parallelism does not have to go through MPI: the XPU kernels support multithreading through \code{KernelAbstractions.jl}, so a single node with many CPU cores runs without \code{MPI.jl}.

Beyond one node, the same code runs under \code{mpiexec} on a cluster (Section~\ref{sec:ui}).
The narrow communication interface keeps both modes on one code path.
The scaling of this design is quantified in Section~\ref{sec:performance}.

\begin{figure}[t]
\begin{lstlisting}
using AEther

γ  = 5 / 3
B0 = 1 / sqrt(4π)

A(x, y, z) = (0, 0, B0 * (cos(4π * x) / 4π +
                          cos(2π * y) / 2π))

w(x, y, z) = (25 / (36π),           # density
              -sin(2π * y),         # v1
              sin(2π * x),          # v2
              0,                    # v3
              5 / (12π) / (γ - 1))  # int. energy

mesh = Mesh(CPU(); size = (256, 256, 1),
            extent = (1, 1, 1),
            cells_per_block = (64, 64, 1),
            nghost = 3)

sim = Simulation(mesh;
                 eos = IdealMHD(γ, 1e-12, 1e-10),
                 reconstruction = WENOZ(),
                 riemann_solver = HLLD(),
                 stepper = RK3(),
                 cfl = 0.4)

set_initial_condition!(w, sim;
                       vector_potential = A)

run!(sim; stop_time = 0.5)
\end{lstlisting}
\caption{Complete \aether\ setup of the Orszag-Tang vortex: an ideal-MHD run on a $256^{2}$ grid with WENO-Z reconstruction, the HLLD Riemann solver, and third-order Runge-Kutta integration.
Loading \code{CUDA.jl} and replacing \code{CPU()} with \code{GPU(CUDABackend())} runs the same script on a GPU.}
\label{fig:ot_script}
\end{figure}

\subsection{User Interface, Problem Setup, and Execution Modes} \label{sec:ui}

The abstractions described above also define the user interface of \aether.
Their practical value is that even a complicated numerical experiment can be expressed concisely in the same vocabulary used to describe its physics and numerical methods.
Traditional simulation codes often distribute problem setup between a parameter file and a problem-generator source file, which must be edited and recompiled for customized experiments.
\aether\ is instead used as a Julia library.
Users define the initial state and custom physics as ordinary functions, assemble a \code{Mesh} and a \code{Simulation} through constructors, and call \code{run!}.
No separate parameter parser, problem-generator registry, or build step stands between the mathematical specification and the code that is executed.

Figure~\ref{fig:ot_script} demonstrates this design with the Orszag--Tang vortex \citep{OrszagTang1979}.
The complete setup and launch contain only 9 lines of Julia code.
The functions \code{w} and \code{A} directly state the primitive variables and magnetic vector potential of the initial condition.
The \code{Mesh} constructor specifies the domain, resolution, block decomposition, and execution backend, while the \code{Simulation} constructor selects the equation of state, reconstruction scheme, Riemann solver, and time integrator.
The final two statements initialize the state and advance it to the requested time.
No prewritten Orszag--Tang problem generator is invoked; all problem-specific information appears in the listing.
Its brevity therefore comes from composing general interfaces rather than hiding configuration elsewhere.

The same problem description operates in three settings.
It can be pasted into the Julia REPL for interactive execution, placed in a Jupyter notebook together with analysis and visualization, or launched as \code{mpiexec -np 8 julia script.jl} under a cluster batch scheduler.
The physical and numerical specification remains unchanged across these modes.
Scaling a prototype into a production run requires changing only the mesh or backend arguments and the launch command, rather than translating the problem into another language or a compiled source module.

Because \aether\ is a library rather than a separate executable, the \code{Simulation} object remains accessible after \code{run!} returns, and its fields are ordinary Julia arrays.
A user can inspect slices, compute diagnostics, and visualize the state in the same process that performed the simulation.
The script that defines the problem therefore also serves as an executable record of the experiment and as the starting point for its analysis.
This continuity removes the translation step of the two-language workflow described in Section~\ref{sec:intro}.

\rev{A script-driven workflow of this kind is not unique to Julia: lower-level C, C++, and Fortran codes achieve it for setup and analysis through Python bindings.
A binding, however, exposes only the hooks and arrays that the compiled side was designed to expose, and user code stays outside the solver's kernels.
Writing the solver natively in the scripting language removes that boundary.
What a user can change is then limited by the problem rather than by the interface.
New physics and numerical methods enter at the same level, and in the same session, as the problem that requires them (Section~\ref{sec:xpu}).}
User-defined boundary conditions attach as Julia functions (Section~\ref{sec:mesh}), source terms use the interface of Section~\ref{sec:sources}, and custom diagnostics enter through the history columns of Section~\ref{sec:io}.
A complete experiment, including its parameters, custom physics, execution settings, and analysis, can therefore remain in one version-controlled Julia file.
The following subsection describes how field and restart output extend this workflow to data stored on disk.

\subsection{Output and Restart} \label{sec:io}

Output is handled by \code{Output} structs attached to the simulation at construction and scheduled inside \code{run!}.
Three writers cover the standard needs.
\code{HistoryOutput} appends volume-integrated diagnostics to a text file written by rank zero: total mass, momentum, and energy, the maximum $|\divB|$ for MHD, and user-defined sum, minimum, or maximum columns.
Conservation and stability can therefore be monitored while a run executes.
\code{FieldOutput} writes selected primitive, conserved, or derived fields, such as vorticity or current density, in single precision by default.
\code{RestartOutput} writes the complete state that continues a run, at working precision.

The \code{Output} checks the simulation time after every cycle and fires when the time exceeds the next scheduled dump.
The history writer can also fire on cycle counts, and the restart writer on elapsed wall-clock time, which protects long runs against batch queue limits.
A forced dump completes every sequence at the stop time, so the final state always reaches disk.

To handle the output efficiently, Field and restart data are written through ADIOS2 \citep{Godoy2020} using its BP5 engine.
ADIOS2 was developed within the United States Exascale Computing Project and handles production output for simulation codes at leadership-class facilities.
Adopting it delegates parallel writing, aggregation, and format evolution to an established library.
A dump stores every variable as one array in global cell space, and ADIOS2 aggregates the writes within each node.
Analysis tools therefore address the domain directly, which is the access pattern uniform-grid analysis needs.

In addition, each file is self-describing. Alongside the arrays it also carries meta data such as mesh-level and version attributes, so a reader reconstructs the grid without access to the generating script.
The same metadata leaves room for growth: for example, the mesh refinement (Section~\ref{sec:mesh}) extends the format with few modifications.
Because ADIOS2 provides Julia and Python bindings, the files open directly in the environments where analysis happens.

\section{Human-AI Development Framework} \label{sec:workflow}

\rev{Building a production simulation code has traditionally taken a research group several years.
Interactive coding agents have begun to compress that timeline.
Operating inside a repository, these tools translate natural-language specifications into multi-file patches and resolve a substantial share of benchmark software issues \citep{Jimenez2024, Yang2024}.
Lower coding cost does not relax the requirements of a scientific simulation code, which still depends on numerical reliability and long-term maintainability.
We built \aether\ to test whether a human-AI collaboration can meet those standards across the full construction of a production code.}

\rev{The main obstacle is the gap between finishing a self-contained task and sustaining long term development across multiple sessions.
Coding agents work well when a task fits inside a single session, such as writing an analysis script or fixing an isolated bug.
In that one-shot regime, the agent's working context and intermediate lessons can be discarded as soon as the patch lands.
A production simulation code cannot be built that way.
Its modules are tightly coupled and must evolve across dozens of sessions and successive model releases without losing their shared design philosophy.}

\rev{Preserving that philosophy runs into several practical problems.
Project knowledge does not carry over automatically: the source code shows what each routine does, but not how modules fit together or why earlier designs were rejected.
Models also differ in how they structure abstractions and write code, so unguided edits gradually pull the solver away from a unified design.
Finally, human judgment from review and debugging stays tied to one session or local device, leaving later sessions or other machines to repeat the same mistakes.}

\rev{We address these problems by designing a closed-loop human-AI framework that governs how agents access project knowledge, how tasks are scoped, and how generated code is verified.
At the file level, modern agent harnesses such as Claude Code and Codex provide flexible hooks for loading repository documents and custom tools into a session \citep{Anthropic2025}.
Whereas those hooks supply general-purpose mechanisms for a single session, our framework builds the surrounding engineering architecture required for a production solver.
Project knowledge, coding standards, and session lessons live in git alongside the Julia source, keeping the memory synchronized across machines and independent of any single vendor tool.
Human developers and coding agents read the same documents and test code interactively through a persistent Julia REPL server.}

\rev{The following subsections detail each layer of this architecture.
Section~\ref{sec:files} traces the intellectual lineage of the memory system and describes each document layer.
Section~\ref{sec:sessions} presents our workflow decomposition and the self-correcting feedback loop that distills human interventions into permanent project rules.
Section~\ref{sec:verification} details the multi-tier physics and software verification suite and the division of testing duties between human and agent.
Section~\ref{sec:record} analyzes the 60-session empirical record to map the current boundaries of AI automation in scientific computing.
Section~\ref{sec:agent-assisted-execution} then shows how end users draw on the same repository memory to configure and launch production simulations.}

\subsection{\rev{Repository-Resident Project Memory System}} \label{sec:files}

\begin{table*}[t]
\centering
\caption{\rev{The five-layer repository memory system that carries project knowledge across sessions, models, and machines.
Every file evolves with the code except the session journal, which is append-only.
The same top-level file answers to the names \code{AGENTS.md} and \code{CLAUDE.md}, the conventions different harnesses search for.}
\label{tab:files}}
\begin{tabular}{@{}lll@{}}
\toprule
\rev{Files} & \rev{Contents} & \rev{Enters an agent's context} \\
\midrule
\rev{\code{AGENTS.md}} & \rev{project map: layout, commands, core rules} & \rev{always, at session start} \\
\rev{\code{.claude/rules/}} & \rev{scoped rulebooks: naming, kernels, tests, docs} & \rev{when editing a file in the rule's scope} \\
\rev{\code{.claude/skills/}} & \rev{procedures: environments, problem runs, cluster jobs} & \rev{on demand, when the task matches} \\
\rev{\code{dev/design-*.md}} & \rev{module designs: decisions, interfaces, rejected alternatives} & \rev{at task start, for the modules touched} \\
\rev{\code{dev/log/}} & \rev{session journal: prompts, interventions, lessons} & \rev{written before every commit; read for audits} \\
\bottomrule
\end{tabular}
\end{table*}

\rev{\aether\ organizes its project memory into the five repository layers listed in Table~\ref{tab:files}.
Keeping all five layers in git makes the memory persistent across sessions, models, and machines, while subjecting every update to the same review as the source code.
Loading the entire documentation tree into every prompt would quickly crowd the context window, so each layer is assigned a specific rule for when it enters a session.
Only the top-level project map loads unconditionally; the remaining layers load on demand when a task touches their scope.}

\rev{This design is an experimental fusion of ideas that have circulated individually across agentic programming and software engineering, adapted here for long-term simulation code development.
Splitting the repository into rules, operational procedures, module designs, and session logs reflects the separation of procedural, semantic, and episodic memory in language-agent research \citep{Shinn2023, Wang2023, Sumers2024}.
The first three layers of Table~\ref{tab:files} draw on the file-loading conventions popularized by vendor guides \citep{Anthropic2025} and the open \code{AGENTS.md} standard.}\footnote{\url{https://agents.md}}
\rev{We fuse those hooks with pre-implementation module contracts inspired by Architecture Decision Records \citep{Nygard2011} to preserve cross-module solver invariants.
An append-only session journal (\code{dev/log/}) then records every human intervention and feeds recurring lessons back into the contracts and rulebooks at commit time (Section~\ref{sec:sessions}).}

\rev{\code{AGENTS.md} is a concise map of the repository.
Loaded at the start of every session, it lists the directory layout, the role of each module, and the commands for building environments and running tests.
It also states the core implementation rules from Section~\ref{sec:software}, such as preferring multiple dispatch over runtime branches and keeping GPU kernels backend-agnostic.
Detailed instructions live in the scoped rulebooks, keeping \code{AGENTS.md} under 150 lines so it occupies minimal context.}

\rev{The rulebooks under \code{.claude/rules/} enforce consistent code style across sessions and models.
Seven files, totaling roughly 300 lines, cover naming and comments, GPU kernels, docstrings, testing, documentation, examples, and interactive REPL usage.
Each file header declares the repository paths it governs, and the harness loads matching rulebooks automatically when the agent edits those paths.
For instance, \code{kernel-rules.md} requires allocation-free and type-stable kernels, bans \code{Float64} literals that would pollute \code{Float32} runs, and favors \code{ifelse} over divergent branches (Section~\ref{sec:xpu}).
Every model that edits \code{src/} works under the same constraints.}

\rev{Skills under \code{.claude/skills/} record how to run the code rather than how to write it.
Each skill documents one operational procedure, such as setting up a scratch environment, running a test problem on CPU or GPU, or launching a multi-node MPI job.
The harness loads a skill only when a task matches its description.
Each guide includes verified throughput and error baselines, saving the agent from spending a session rediscovering launch flags by trial and error.
Human users read the same files as quick-start guides (Section~\ref{sec:agent-assisted-execution}).}

\rev{The design documents \code{dev/design-*.md} consist of a cross-module overview and 17 module files that separate architectural choices from coding.
Each file is drafted during a dedicated design session (Section~\ref{sec:sessions}) before implementation begins, then updated to match the finished module.
Beyond public signatures and array layouts, each document records the alternatives that were considered and rejected.
Recording rejected ideas prevents later sessions from re-proposing dead ends, such as dimension-sequential ghost passes that serialize three MPI rounds per stage.
A new session reads only the overview and the design documents of the modules it will edit.}

\rev{The session journal under \code{dev/log/} is the only append-only layer in Table~\ref{tab:files}.
Whereas the other four layers track the current state of the code, \code{dev/log/} adds one permanent entry per working session across the 60 logged sessions.
Each entry records the user prompts and human interventions, preserving the reasoning behind superseded designs much as git history preserves old code.
The next subsection describes how sessions are scoped and how a git hook enforces this journal.}

\subsection{\rev{Session Workflow and the Self-Correcting Feedback Loop}} \label{sec:sessions}

\rev{File scoping limits how much text enters a session at startup; task decomposition limits how much context accumulates while the agent works.
Even when a model advertises a $10^{6}$-token window, retrieval accuracy drops well before that ceiling as early constraints get buried in a long transcript \citep{Liu2024}.
In our work on \aether, agent sessions became noticeably less reliable past roughly $2\times10^{5}$ tokens, particularly after debugging detours left failed attempts in the conversation history.
Every task is therefore sized to finish within a single session below that threshold.}

\rev{We also separate design from implementation by splitting each module across two sessions.
The first session is an interactive design discussion in which no source code is written.
The human and the agent work through data layouts, function signatures, and coupling to existing modules, checking candidates against the constraints already in \code{dev/}.
The session ends by writing the module's design document, which serves as the contract for implementation.
If a task would still require an architectural decision mid-way through coding, we split it into smaller tasks before moving on.}

\rev{Implementation runs in a new session with a clean context window.
The agent loads \code{AGENTS.md}, the path-matched rulebooks, and the module design document, then proposes an implementation plan for human approval.
Once approved, the agent writes the code according to the design document and checks individual functions in a persistent Julia REPL (Section~\ref{sec:ui}) as it goes.
If an unforeseen issue requires departing from the design document, the agent will stop and ask rather than improvise.
Before presenting its diff, the agent runs the CPU test suite, checks new docstrings, and audits its diff against the rulebooks.}

\rev{An implementation session often misses flaws in code it just wrote, so we review nontrivial diffs in a separate session, frequently using a second model family.
The reviewing agent needs only the diff, the design document, and the rulebooks, so this cross-model audit uses minimal context.
Throughout this process, the human owns the architecture, physical validation, rulebooks, and commits; the agent writes the module code, test scripts, and docstrings.}

\label{sec:hooks}\rev{A versioned git pre-commit hook (\code{.githooks/pre-commit}) ties the session workflow back to the project memory.
The hook rejects any commit touching code, tests, or documentation unless a new entry in \code{dev/log/} is staged in the same commit.
This check runs inside git rather than a specific harness, so it applies equally to Claude Code, Codex, and the human developer.
Each log entry follows a template recording the tool, exact model version and reasoning effort, quoted user prompts, files produced, and every human intervention.
The commit message ends with a matching \code{Written-by:} trailer naming the model version and effort.}

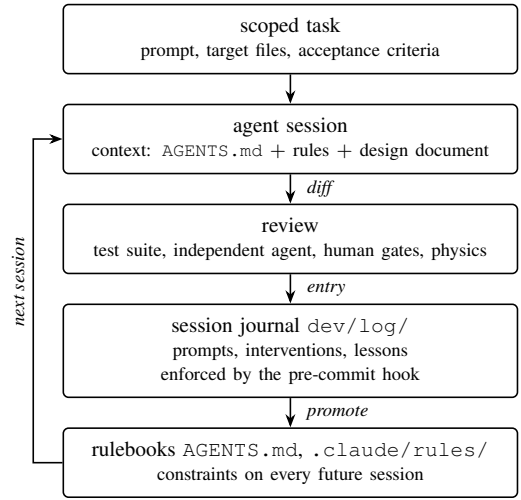
\begin{figure}[t]
\centering
\begin{tikzpicture}[
    font=\footnotesize,
    box/.style={draw, semithick, rounded corners=2pt, align=center,
                text width=0.68\columnwidth, inner ysep=5pt},
    elab/.style={font=\scriptsize\itshape},
    arr/.style={-{Stealth[length=5pt]}, semithick},
    node distance=11pt
]
\node[box] (task)   {scoped task\\{\scriptsize prompt, target files, acceptance criteria}};
\node[box, below=of task] (agent)  {agent session\\{\scriptsize context: \code{AGENTS.md} $+$ rules $+$ design document}};
\node[box, below=of agent] (review) {review\\{\scriptsize test suite, independent agent, human gates, physics}};
\node[box, below=of review] (log)   {session journal \code{dev/log/}\\{\scriptsize prompts, interventions, lessons}\\{\scriptsize enforced by the pre-commit hook}};
\node[box, below=of log] (rules)   {rulebooks \code{AGENTS.md}, \code{.claude/rules/}\\{\scriptsize constraints on every future session}};
\draw[arr] (task)   -- (agent);
\draw[arr] (agent)  -- node[elab, right=3pt] {diff} (review);
\draw[arr] (review) -- node[elab, right=3pt] {entry} (log);
\draw[arr] (log)    -- node[elab, right=3pt] {promote} (rules);
\draw[arr] (rules.west) -- ++(-11pt,0) |-
    node[elab, rotate=90, anchor=south, pos=0.28] {next session} (agent.west);
\end{tikzpicture}
\caption{\rev{The feedback loop of the development workflow.
A scoped task enters an agent session whose context holds the project map, the matching rulebooks, and the design document (Section~\ref{sec:files}).
Its diff passes the review and verification gates (Sections~\ref{sec:sessions} and \ref{sec:verification}).
The git pre-commit hook then forces a session-journal entry, and lessons recorded there are promoted, with human approval, into the rulebooks that constrain every following session.}}
\label{fig:loop}
\end{figure}

\rev{Writing the log entry closes the feedback loop shown in Figure~\ref{fig:loop}.
At the end of a session, the agent reviews the human corrections recorded in the log and proposes memory updates so the same intervention is not needed twice.
To keep the global instruction files compact, module-specific updates go into \code{dev/design-*.md} and one-off fixes stay in the log as candidates if they recur.
Repeated cross-cutting mistakes become rules in \code{AGENTS.md} or \code{.claude/rules/}.
Approved updates ship in the same commit as the code and log entry, working alongside the verification suite described next.}

\subsection{\rev{Multi-Tier Physics and Software Verification}} \label{sec:verification}

\rev{In general software projects, a passing test suite is often enough to accept an agent's patch.
In a PDE solver, an agent can make a test pass by widening an error tolerance while breaking conservation, symmetry, or convergence order.
Our workflow therefore separates writing test code from choosing what the test checks.
The agent writes the test scripts and doctests during implementation, whereas the human specifies the target identities, approves every tolerance, and inspects the physical solutions.
The testing rulebook (\code{testing-rules.md}) explicitly bans guessed error bounds.
For each convergence test, we first measure the error across resolutions and confirm that the rate matches the scheme's formal order \citep{Oberkampf2010}.
We then lock the regression threshold just below that measured rate.}

\rev{\aether\ verifies changes at three levels: function-level unit tests, pipeline integration tests, and full physics problems, totaling roughly 3,200 automated assertions.
When the core modules were first being written, we used 18 per-module unit-test files (about 3,300 lines) to test structs and helpers as they came online.
Once the full solver pipeline was assembled, we retired scaffolding tests that only checked internal boilerplate and consolidated the mathematical unit tests into the first tier.
This pruning cut the CPU test runtime to about 70 seconds, fast enough for an agent to run via Julia's \code{Pkg.test()} runner before every commit.
The same suite runs on every push and pull request through GitHub Actions continuous integration with Codecov coverage tracking.}

\rev{Unit tests in the first tier check that individual functions and kernels behave as intended in isolation before they are wired into a simulation.
Here, \code{Aqua.jl} checks package hygiene and method ambiguities, while 88 \code{jldoctest} blocks in \code{src/} test public constructors, thermodynamic conversions, and coordinate metrics on hand-built inputs.
Additional unit tests in \code{test/software.jl} and \code{test/output.jl} pin algebraic identities and past edge-case bugs.
For example, they verify that \code{LHLLC} reduces to \code{HLLC} at unit Mach scaling and that \code{LHLLC} and \code{LHLLD} preserve reflection symmetry across all directions.}

\rev{Once these building blocks are wired into a \code{Simulation}, the second tier verifies that the coupled modules interact properly during time-step advancement and agree across execution backends.
Running on small multidimensional grids with simple operation, these pipeline tests check that ghost-cell exchanges, source-term kicks, and \code{IMEX2P} stages execute in the right order and leave a uniform state bitwise unchanged.
They also confirm that mass, momentum, energy, and a divergence-free magnetic field are preserved on both CPU and GPU in \code{Float32} and \code{Float64}.
Under the environments of 4- and 8-rank MPI, the same tier verifies Morton block distribution and confirms that ghost exchanges, ADIOS2 field dumps, and checkpoint restarts match the serial run's result.}

\rev{The third tier tests the physical accuracy of the solver on practical problems.
It runs more than 20 one-, two-, and three-dimensional hydrodynamic, MHD, curvilinear, and multifluid dust test problems against exact solutions or reference baselines, matching the validation problems in Section~\ref{sec:tests}.
As a result, this three-tier design provides comprehensive testing for the package and isolates the source of a bug whenever a new change breaks a test.}

\subsection{\rev{Development Record and Current Boundaries of Coding Agents}} \label{sec:record}

\begin{table*}[t]
\centering
\caption{\rev{Summary of the 60 development sessions recorded in \code{dev/log/} between July 8 and August 30, 2026.
The upper block partitions all 60 sessions by their outcome at the human review gate.
The middle block classifies the 40 sessions that required human intervention or reversion by their primary technical area.
The lower block counts the sessions that updated persistent project memory in the same commit.}
\label{tab:record}}
\begin{tabular}{@{}lrr@{}}
\toprule
\rev{Category} & \rev{Sessions} & \rev{Share (\%)} \\
\midrule
\multicolumn{3}{@{}l}{\rev{\emph{Outcome at the human review gate (60 sessions, 100.0\%)}}} \\
\rev{Accepted as delivered without correction} & \rev{20} & \rev{33.3} \\
\rev{Steered or overruled at the design stage} & \rev{12} & \rev{20.0} \\
\rev{Corrected or redirected during code review} & \rev{27} & \rev{45.0} \\
\rev{Merged and later reverted wholesale} & \rev{1} & \rev{1.7} \\
\midrule
\multicolumn{3}{@{}l}{\rev{\emph{Primary area of human intervention (40 sessions, 66.7\%)}}} \\
\rev{Software architecture and abstraction discipline} & \rev{13} & \rev{21.7} \\
\rev{Physics formulation and numerical schemes} & \rev{12} & \rev{20.0} \\
\rev{GPU portability and parallel scaling} & \rev{6} & \rev{10.0} \\
\rev{Naming, documentation, and workflow scope} & \rev{9} & \rev{15.0} \\
\midrule
\multicolumn{3}{@{}l}{\rev{\emph{Project memory updates (38 unique sessions, 63.3\%)}}} \\
\rev{Rule promotions (\code{AGENTS.md}, \code{.claude/rules/})} & \rev{13} & \rev{21.7} \\
\rev{Design document updates (\code{dev/design-*.md})} & \rev{31} & \rev{51.7} \\
\bottomrule
\end{tabular}
\end{table*}

\rev{Building \aether\ in about 2 months across 60 logged sessions (Table~\ref{tab:record}) shows how coding agents change the day-to-day work of computational science.
A central question in the community is whether coding agents can scale beyond small scripts to build a full simulation code.
Our record shows that they can already carry the implementation of a production solver from scratch.
Once a module's equations, data layout, and interfaces were fixed in a design document, the agent delivered working, test-passing code on the first try in one-third of all sessions.
Even in the remaining two-thirds, the agent still automated most of the coding, and only one session had to be reverted wholesale.
Rather than fixing basic syntax, the human stepped in to guide the design philosophy, make key physical choices, or keep the agent from getting hung up on minor details.
The agent therefore does not replace the domain expert, but shifts human effort from writing boilerplate to high-level direction.}

\rev{Examining the 40 sessions that required human intervention reveals two limits of current coding agents.
First, language models are trained mostly on general-purpose software and tend to over-complicate scientific code.
Left to themselves, agents introduced heavy object-oriented frameworks in a multiple-dispatch language, such as proposing a dynamic task scheduler where a fixed sequence of time-step stages sufficed.
They also attempted premature GPU optimizations, such as fusing multi-species dust loops in a way that obscured how the code maps to the equations without speeding up memory-bound GPU kernels.
Second, agents tend to focus on one dimension of a task at a time, such as making a test pass, while missing couplings across code, numerics, and physics.
In one session, a port of a low-Mach Riemann solver \citep{Minoshima2021} passed standard shock tests even though a missing pressure term canceled out the low-Mach momentum fix.
Catching that defect required checking the discrete algebra line by line against the paper.
Similarly, when a 3D radiative MHD turbulence run blew up on multiple GPUs, the agent searched crash logs for a software bug rather than recognizing that high-order overshoots were driving a magnetic runaway.
Diagnosing that coupling and switching to implicit cooling with monotone reconstruction required a physicist.}

\rev{The feedback loop in Figure~\ref{fig:loop} turns these human interventions into cumulative project memory rather than repeated effort.
In 38 of the 60 sessions, review corrections were distilled into updated design documents or permanent repository rules (Table~\ref{tab:record}).
For example, after early sessions repeatedly placed slow input checks inside the inner time-stepping loop, adding one rule restricting validation to setup stopped that mistake permanently.
While no rulebook can replace cross-disciplinary physical judgment, encoding architectural and numerical boundaries steadily narrows the space of software errors as the codebase grows.
This feedback loop between human steering and the coding agent improves each subsequent session, freeing the scientist to focus on physics while leaving a documented repository for running simulations.}

\subsection{\rev{Agent-Assisted Simulation with Natural Language}} \label{sec:agent-assisted-execution}

\rev{The same markdown memory that guides code development also lets users set up and run simulations with an agent.
When a user asks an agent to configure a problem, the agent reads \code{AGENTS.md}, the relevant design documents, and the \code{run-problem} skill.
That skill specifies how to build an environment, structure a run script, and compare outputs against reference data.
Example scripts and notebooks in the repository then serve as physical templates that the agent adapts to the user's equations and geometry.}

\rev{In a typical session, the user specifies the fluid model, domain geometry, boundary conditions, diagnostics, and target device in natural language.
The agent writes the corresponding \code{Mesh}, \code{Simulation}, and \code{Output} setup (Section~\ref{sec:software}) and runs a low-resolution test case first.
Checking this short run against the test suite catches configuration or decomposition mistakes before cluster allocation time is spent.}

\rev{For production runs, cluster skills supply the job-scheduler directives, rank-to-GPU binding flags, and MPI environment variables required on multi-node machines.
The agent can generate the batch script and launch the full calculation without altering the problem definition in the Julia script.
The validation problems in Section~\ref{sec:tests} both verify the numerical methods of \aether\ and provide the reference baselines used by this operational workflow.}

\section{Numerical Tests} \label{sec:tests}

To verify that \aether\ produces correct results and to catch regressions during development, we maintain a suite of standard test problems \citep[e.g.,][]{Fryxell2000, Stone2008, Stone2020, Stone2026}.
When an exact or semi-analytic solution is available, we use it to measure the numerical error directly.
For problems without such a solution, we compare the numerical result with a higher-resolution reference calculation that has been checked for convergence, or verify a property that must hold exactly, such as symmetry.
The examples presented in this section form a representative subset chosen to exercise the code's main physical models, numerical methods, and mesh geometries.
The full regression suite under \code{test/} extends this coverage with additional tests of accuracy, robustness, boundary conditions, and distributed execution.

All tests presented here can be run through the public interface described in Section~\ref{sec:ui}.
Jupyter notebooks that reproduce most of the figures are distributed with the code.\footnote{\url{https://github.com/MHDFlows/Aether-Example}}

\subsection{Sod Shock Tube} \label{sec:sod}

\begin{figure*}[t]
\centering
\includegraphics[width=\textwidth]{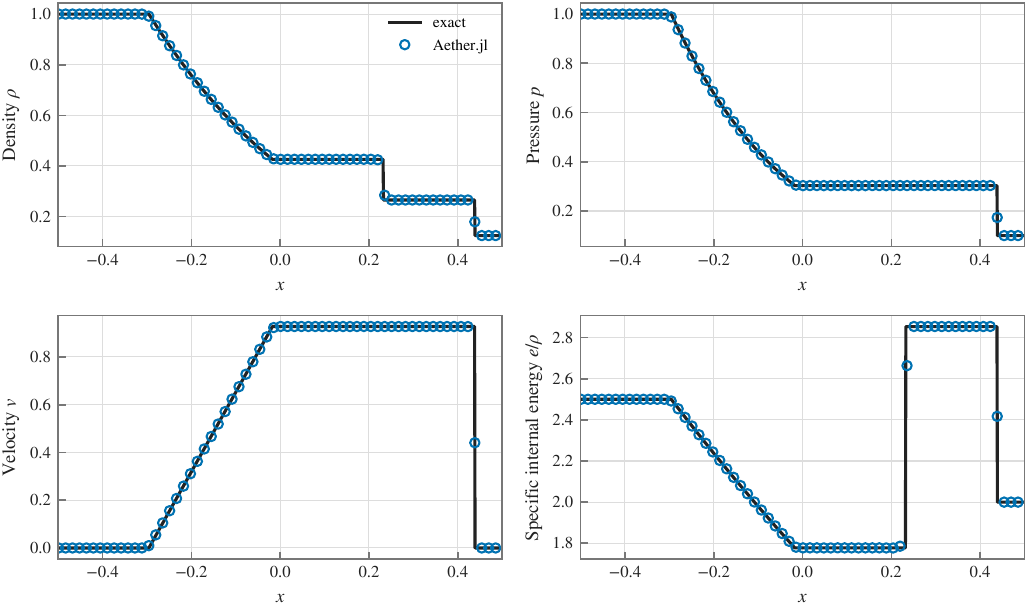}
\caption{Sod shock tube at $t = 0.25$: density, pressure, velocity, and specific internal energy (every eighth cell shown).}
\label{fig:sod}
\end{figure*}

We begin with the shock tube of \citet{Sod1978}, the standard first test of a shock-capturing code because its exact solution contains all three hydrodynamic wave families at moderate strength.
Before any finer property can be measured, a Godunov scheme must place these nonlinear waves at the correct positions with the correct amplitudes.
The initial condition separates two resting states at $x = 0$, with $(\rho, p) = (1, 1)$ on the left and $(0.125, 0.1)$ on the right, and $\gamma = 1.4$.
We use 512 cells on $x \in [-0.5, 0.5]$ with PPM reconstruction, the HLLC solver, RK3, and a Courant factor of 0.8.

Figure~\ref{fig:sod} compares the numerical solution at $t = 0.25$ with the exact one.
The initial discontinuity breaks up into a left-moving rarefaction fan, a contact discontinuity, and a right-moving shock.
The numerical profiles reproduce the position and strength of all three waves.
Both discontinuities remain sharp, spreading over a few cells, and no spurious oscillations develop behind either front.
The specific internal energy, the most sensitive of the four quantities shown, deviates only in isolated cells inside the two fronts, the usual footprint of captured discontinuities.

\subsection{Shu-Osher and Brio-Wu Shock Tubes} \label{sec:shocktubes}

\begin{figure*}[t]
\centering
\includegraphics[width=\textwidth]{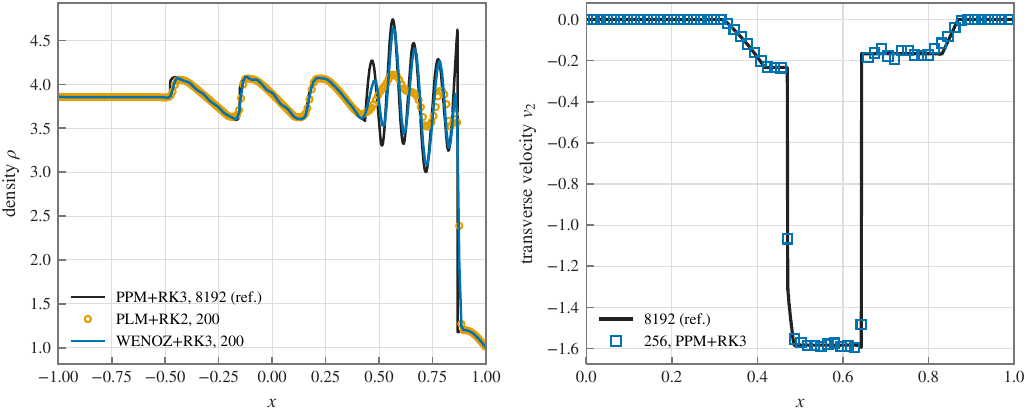}
\caption{Left: density in the Shu-Osher problem at $t = 0.47$.
Right: transverse velocity $v_{2}$ in the Brio-Wu MHD shock tube at $t = 0.1$ (every fourth cell shown).}
\label{fig:shocktubes}
\end{figure*}

We next consider two harder one-dimensional problems, collected in Figure~\ref{fig:shocktubes}, which probe what the Sod tube cannot: the fidelity of the reconstruction behind a shock, and the wave families specific to MHD.

The test of \citet{ShuOsher1989} sends a Mach~3 shock, initially at $x = -0.8$, into the sinusoidally perturbed density field $\rho = 1 + 0.2\sin(5\pi x)$.
The shock compresses the incoming entropy waves into short-wavelength oscillations that numerical diffusion damps, so the test measures how much of this structure a scheme retains at fixed resolution.
The left panel of Figure~\ref{fig:shocktubes} compares two 200-cell runs with the HLLC solver against a reference solution computed with PPM on 8192 cells.
Fifth-order WENO-Z reconstruction with RK3 recovers nearly the full amplitude of the post-shock oscillations, while second-order PLM with RK2 visibly clips them \citep[cf.][]{Stone2020}.
This difference motivates the higher-order reconstructions of Section~\ref{sec:fluxes} for flows that carry fine structure through shocks.

The right panel shows the MHD shock tube of \citet{BrioWu1988}.
It extends the Sod initial condition with $\gamma = 2$, a normal field $B_{1} = 0.75$, and a transverse field that reverses from $+1$ to $-1$ across the interface.
Its breakup produces wave structures specific to MHD, including a compound wave.
Because no exact solution exists, the reference is again a self-converged run on 8192 cells.
We plot the transverse velocity $v_{2}$, which vanishes initially and therefore responds to every wave in the fan.
With 256 cells, PPM reconstruction, the HLLD solver, and RK3, the numerical profile tracks the reference through all fronts and resolves each discontinuity with one to two interior points.

\subsection{Circularly Polarized Alfv\'en Wave} \label{sec:cpaw}

\begin{figure}[t]
\centering
\includegraphics[width=\columnwidth]{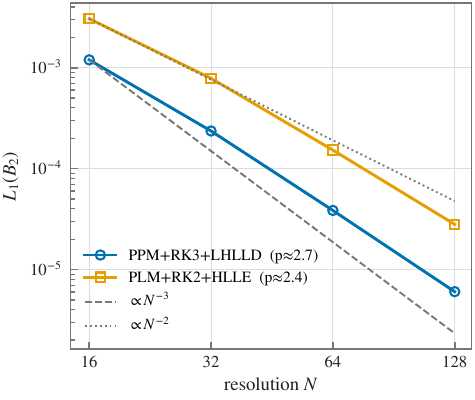}
\caption{$L_{1}$ error of $B_{2}$ for the circularly polarized Alfv\'en wave after one crossing of the periodic domain, as a function of resolution $N$.
The legend quotes the convergence order measured over the finest refinement interval; dashed and dotted guides show slopes of $-3$ and $-2$.}
\label{fig:cpaw}
\end{figure}

Shock tubes confirm wave structure but provide no convergence rate, because discontinuities limit every scheme to first-order accuracy locally.
We therefore measure the order of accuracy with the circularly polarized Alfv\'en wave, the standard smooth test for constrained-transport codes \citep{Toth2000, GardinerStone2008}.
Because the wave propagates without change of shape as an exact nonlinear solution of ideal MHD, the initial data provide the reference at any later time.

The wave travels along $x$ through a uniform isothermal background with $\rho = 1$, $c_{\rm s} = 1$, and $B_{1} = 1$, so the Alfv\'en speed is unity.
Transverse fields $B_{2} = 0.1\sin(2\pi x)$ and $B_{3} = 0.1\cos(2\pi x)$, with matching velocities $v_{2,3} = -B_{2,3}$, place one wavelength across the periodic unit cube.
We evolve grids of $2N \times N \times N$ cells, with $N$ from 16 to 128, for one crossing time at a Courant factor of 0.3.
The $L_{1}$ error of $B_{2}$ against the initial state then measures the convergence.

Figure~\ref{fig:cpaw} shows the errors for two scheme combinations.
PPM reconstruction with the LHLLD solver and RK3 converges at a measured order of 2.7 over the finest refinement interval.
PLM with HLLE and RK2 converges at order 2.4, with errors 2 to 5 times larger at every resolution.
Limiter activity near the wave extrema holds the parabolic scheme below its formal third order, as is common for limited reconstructions on this test.

\subsection{Liska-Wendroff Implosion} \label{sec:implosion}

\begin{figure}[t]
\centering
\includegraphics[width=\columnwidth]{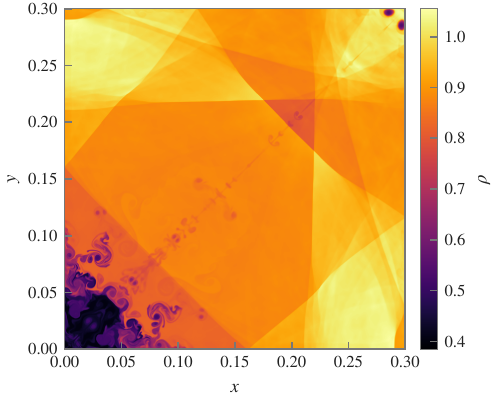}
\caption{Density in the Liska-Wendroff implosion at $t = 2.5$ on a $1024^{2}$ grid.}
\label{fig:implosion}
\end{figure}

Multidimensional shock reflections expose a property that one-dimensional tests cannot: whether the discrete update treats the two coordinate directions identically.
The implosion problem of \citet{LiskaWendroff2003} makes any violation visible.
A low-pressure wedge in the corner of a reflecting box launches a converging shock, and repeated reflections then drive a jet along the diagonal.
Because the initial state is mirror-symmetric about that diagonal, any asymmetry in the update deflects or destroys the jet, a failure documented for schemes that break the $x$-$y$ symmetry.

The domain $[0, 0.3]^{2}$ initially holds $(\rho, p) = (0.125, 0.14)$ below the line $x + y = 0.15$ and $(1, 1)$ above it, with $\gamma = 1.4$ and reflecting walls.
We evolve a $1024^{2}$ grid with PPM, HLLC, and RK3 at a Courant factor of 0.4 to $t = 2.5$, roughly ten acoustic crossings of the box.

Figure~\ref{fig:implosion} shows the resulting density.
The jet propagates cleanly along the diagonal, and the fine vortical structures on either side mirror each other.
The symmetry holds beyond visual inspection: the density field at $t = 2.5$ equals its transpose in every floating-point bit, so the measured asymmetry is exactly zero.
This result confirms that reconstruction, flux evaluation, and the Runge-Kutta combinations evaluate the two directions with identical arithmetic over thousands of cycles.

\subsection{Orszag-Tang Vortex} \label{sec:ot}

\begin{figure*}[t]
\centering
\includegraphics[width=\textwidth]{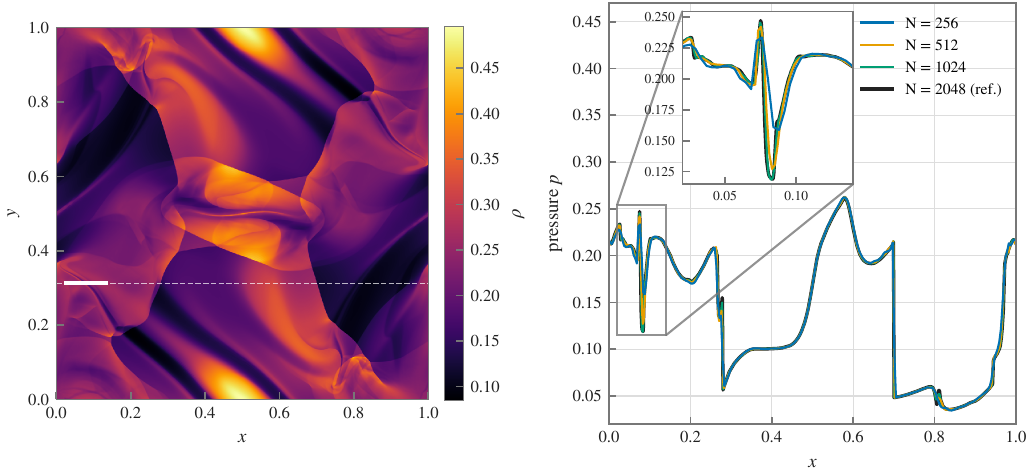}
\caption{Orszag-Tang vortex at $t = 0.5$.
Left: density on the $1024^{2}$ grid; the dashed line marks the cut at $y = 0.3125$, and the solid segment marks the interval magnified in the inset.
Right: pressure along the cut for resolutions of $256^{2}$ to $2048^{2}$; the inset magnifies the marked interval.}
\label{fig:ot}
\end{figure*}

The vortex of \citet{OrszagTang1979} remains the most widely compared multidimensional MHD test, because its smooth initial data steepen into a reproducible network of interacting shocks and current sheets.
The problem already appeared in Section~\ref{sec:ui} as the complete setup script of Figure~\ref{fig:ot_script}; the runs here use the same script with different resolution and scheme settings.
On the periodic unit square, the initial state has $\rho = 25/(36\pi)$, $p = 5/(12\pi)$, $\gamma = 5/3$, and velocity $(-\sin 2\pi y, \sin 2\pi x)$.
The magnetic field derives from the vector potential $A_{3} = B_{0}\left[\cos(4\pi x)/(4\pi) + \cos(2\pi y)/(2\pi)\right]$ with $B_{0} = 1/\sqrt{4\pi}$, so the field starts divergence-free on the faces.
We run resolutions of $256^{2}$ through $2048^{2}$ with PPM5, the LHLLD solver, and RK3 at a Courant factor of 0.4, stopping at $t = 0.5$.

The density map in Figure~\ref{fig:ot} reproduces the familiar morphology of the test at this time, including the central current sheet, the surrounding shock fronts, and the compressed filaments between them \citep[cf.][]{Stone2008, Stone2020}.
For a quantitative comparison, the right panel cuts the pressure along $y = 0.3125$, a line commonly used in cross-code studies.
Away from the sharpest features the four resolutions already coincide, so the inset magnifies $x \in [0.02, 0.14]$, where the narrow spike and dip differ most.
The coarsest run smooths both features, and each refinement roughly halves the remaining deviation until the $1024^{2}$ profile lies on the reference.
All fronts remain free of oscillations.

\subsection{Sedov-Taylor Blast Wave} \label{sec:sedov}

\begin{figure*}[t]
\centering
\includegraphics[width=\textwidth]{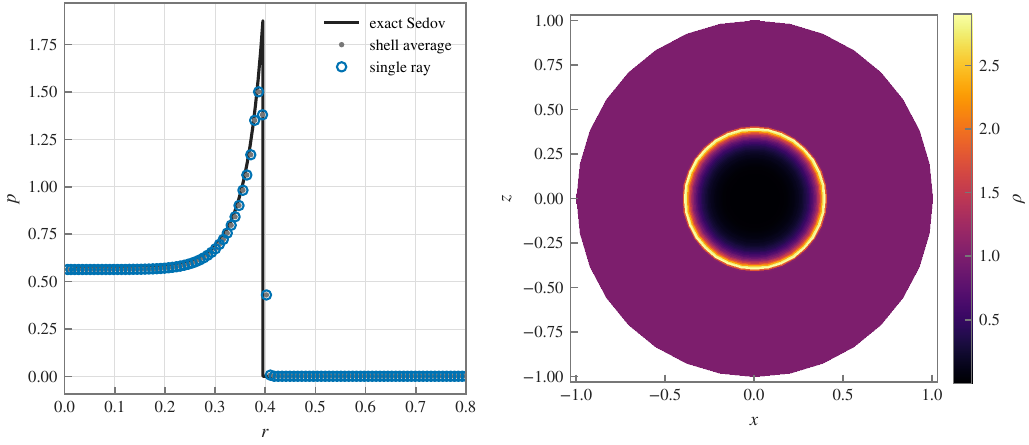}
\caption{Sedov-Taylor blast wave at $t = 0.1$ on the spherical-polar mesh.
Left: pressure against radius; gray points give the shell averages, open circles a single $(\theta, \phi)$ ray, and the black line the self-similar solution for the deposited energy.
Right: density in the meridional plane, mapped to Cartesian coordinates; the grid axis runs vertically through the panel.}
\label{fig:sedov}
\end{figure*}

A point explosion into a cold uniform medium follows the self-similar solution of \citet{Taylor1950} and \citet{Sedov1959}, which fixes the shock radius and the interior profiles at every time.
We run this problem on a spherical-polar mesh, where it exercises the geometric source terms, the exact metric factors, and the polar-axis boundary of Section~\ref{sec:mesh}.
Because the exact flow is spherically symmetric, any angular structure that develops measures the numerical error directly.

The mesh covers the full sphere $r \in (0, 1)$ with $(N_{r}, N_{\theta}, N_{\phi}) = (128, 16, 32)$ cells.
The ambient gas has $\rho = 1$ and $p = 10^{-3}$ with $\gamma = 5/3$, and the blast deposits internal energy uniformly inside $r_{0} = 0.02$.
Because only the two innermost radial shells have centroids inside $r_{0}$, the discrete initial condition holds $E = 0.477$ rather than the nominal unit energy; the analytic profile uses this deposited value.
We combine positivity-preserving WENO-Z reconstruction, HLLC, and RK3 at a Courant factor of 0.3.
The boundaries pair a reflecting inner radius with outflow at $r = 1$ and the polar condition at the axis, and the run stops at $t = 0.1$.

The left panel of Figure~\ref{fig:sedov} compares the pressure with the self-similar solution.
Solid-angle-weighted shell averages over all 512 angular zones coincide with a single sampled ray to the width of the markers.
At each radius, the pressure in every angular zone matches the shell mean to a relative deviation below $3 \times 10^{-13}$, so the blast remains spherically symmetric to near machine precision.
The shock front sits at the predicted radius, while the finite radial resolution smooths the pressure spike behind it and lowers the sampled peak.
The meridional density slice in the right panel shows the same result geometrically: a thin, round shell that crosses the grid axis at the top and bottom of the panel without distortion.

\subsection{Taylor-Green Vortex at Low Mach Numbers} \label{sec:tgv}

\begin{figure}[t]
\centering
\includegraphics[width=\columnwidth]{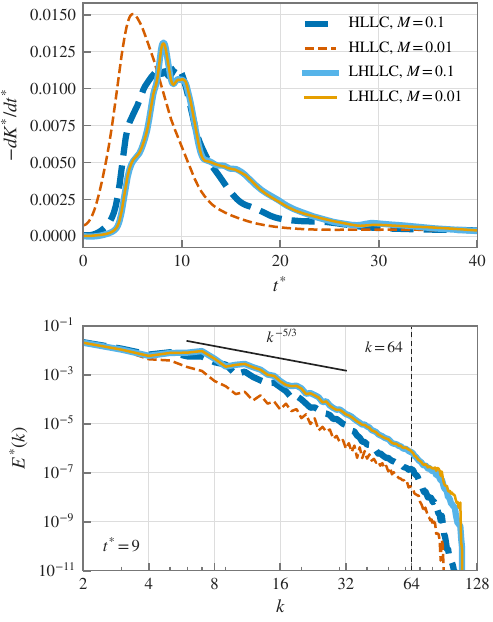}
\caption{\rev{Top:} Kinetic-energy dissipation rate of the three-dimensional Taylor-Green vortex in scaled units, for HLLC (dashed, dark tones) and LHLLC (solid, light tones) at maximum initial Mach numbers of 0.1 (thick blue) and 0.01 (thin orange).
The thin LHLLC curve lies inside the thick one.
\rev{Bottom: Scaled kinetic-energy spectrum $E^{*}(k) = E(k)/V_{0}^{2}$ at $t^{*} = 9$, with a reference $k^{-5/3}$ slope and the Nyquist wavenumber $k = 64$ marked by the vertical dashed line.}}
\label{fig:tgv}
\end{figure}

The tests above exercise the solvers at Mach numbers of order unity or above, yet much astrophysical dynamics is subsonic.
Turbulence in the solar wind and the magnetic reconnection embedded in it evolve at small fractions of the sound speed \citep{BrunoCarbone2013, ZweibelYamada2009}.
In planet-forming disks, the instabilities that concentrate solids into planetesimals grow from velocities of a few percent of the sound speed \citep{YoudinGoodman2005}.
Standard upwind solvers lose accuracy in this regime: their pressure dissipation stays proportional to the sound speed, while the physical pressure fluctuations decrease as the square of the Mach number \citep{GuillardViozat1999}.
These solvers therefore damp subsonic motions that the grid resolves.
To verify that LHLLC removes this failure, we measure the numerical dissipation of a decaying vortex at two Mach numbers, following \citet{Barsukow2017}.

The vortex of \citet{TaylorGreen1937} starts as a single large-scale mode in a periodic cube and decays into progressively smaller vortices.
Because the setup carries no explicit viscosity, the decay of the mean kinetic energy $K$ measures the numerical dissipation alone.
The initial state has uniform density, velocity $\mathbf{v} = V_{0}\left(\sin x \cos y \cos z,\, -\cos x \sin y \cos z,\, 0\right)$, and a uniform pressure that sets the maximum Mach number $M = V_{0}/c_{\rm s}$, plus the pressure perturbation of order $\rho V_{0}^{2}$ that accompanies the vortex.
When time is scaled by the vortex turnover and kinetic energy by $V_{0}^{2}$, runs at different Mach numbers approach the same incompressible limit, so the scaled dissipation rate should not depend on $M$.
We run $M = 0.1$ and $M = 0.01$ on a $128^{3}$ grid with PLM reconstruction and RK2, once with HLLC and once with LHLLC.

\rev{The top panel of} Figure~\ref{fig:tgv} compares the dissipation rates of the four runs.
The two LHLLC curves coincide, agreeing to within 1\% of the peak dissipation rate over the whole run, so the numerical dissipation is independent of the Mach number.
The HLLC curves instead separate: at $M = 0.01$ the peak dissipation rate is 32\% higher than at $M = 0.1$ and occurs at half the scaled time.
This contrast reproduces the behavior of the standard and low-Mach Roe fluxes in Figure 7 of \citet{Barsukow2017}.
\rev{Following \citet{Grinstein2019}, the bottom panel compares the scaled kinetic-energy spectra $E^{*}(k)$ at $t^{*} = 9$, where most of our runs reach peak dissipation.
The two LHLLC spectra again lie on top of each other across resolved wavenumbers ($k \le 64$) and show a tendency toward a $k^{-5/3}$ slope at intermediate scales.
In contrast, HLLC damps small-scale motions in a Mach-dependent way, depleting the spectrum across $k \gtrsim 5$ by nearly an order of magnitude at $M = 0.01$.}
We therefore recommend the low-dissipation solvers for subsonic applications.

\subsection{Multifluid Dust} \label{sec:dusttests}

The remaining tests validate the dust module, which couples an arbitrary number of pressureless fluids to the gas through stiff drag (Section~\ref{sec:dust}).
Its solver relies on algorithms the fluid tests above do not exercise: the pressureless Riemann solver, the closed-form implicit drag stage, and the \code{IMEX2P} integrator of Section~\ref{sec:time}.
We adopt three one-dimensional problems from the benchmark set of \citet{BenitezLlambay2019}, which probe the relevant regimes in turn.
A linear wave tests the accuracy of the coupled oscillation, a nonlinear steady shock tests several species with distinct coupling strengths, and a uniform relaxation problem tests the drift equilibrium under external forcing.
All three problems run isothermal with \code{IMEX2P}.

\subsubsection{Damped Dusty Sound Wave} \label{sec:dustywave}

\begin{figure}[t]
\centering
\includegraphics[width=\columnwidth]{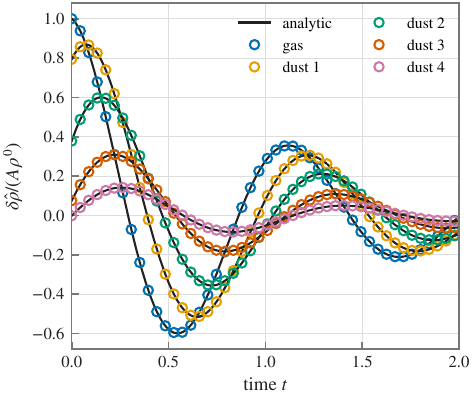}
\caption{Damped dusty sound wave with four dust species: normalized density perturbations at a fixed cell.
Markers subsample the numerical record; black lines show the analytic eigenmode evolution.}
\label{fig:dustywave}
\end{figure}

Sound waves in a dusty medium damp and disperse, because drag exchanges momentum between the phases on each stopping time.
For small amplitudes the coupled equations linearize exactly, so the full time history of every fluid follows from a single complex eigenmode.
This property makes the problem a sharp accuracy test for the coupled gas-dust update.

We use $c_{\rm s} = 1$, $\rho_{\rm g} = 1$, and four dust species with dust-to-gas ratios $\varepsilon = (0.100, 0.233, 0.367, 0.500)$ and stopping times spaced logarithmically from 0.1 to 1.
Linearizing the ten coupled equations at wavenumber $k = 2\pi$ and selecting the least-damped oscillatory mode gives the eigenvalue $\lambda = -0.912 \pm 5.494\,i$.
We initialize all ten fields from the corresponding eigenvector at relative amplitude $10^{-4}$ on 128 cells.
The run then follows the wave for two time units, about 1.7 oscillation periods and 1.8 damping times.

Figure~\ref{fig:dustywave} tracks the normalized density perturbation of every fluid at a fixed cell.
The gas and all four dust species follow the analytic mode in phase and amplitude through the full decay.
This agreement verifies the drag coupling in both directions together with the pressureless advection of each species.

\subsubsection{Steady Dusty Shock} \label{sec:dustyshock}

\begin{figure}[t]
\centering
\includegraphics[width=\columnwidth]{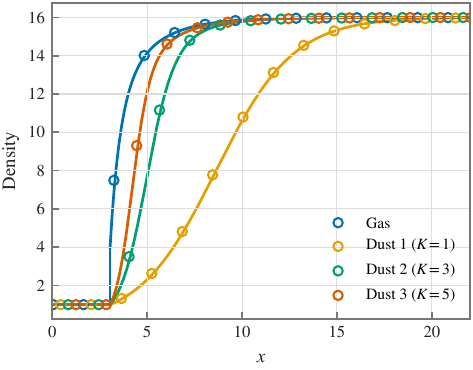}
\caption{Steady isothermal dusty shock with three dust species.
Open markers show the stationary \aether\ profiles; solid lines the semi-analytic reference.}
\label{fig:dustyshock}
\end{figure}

Behind a shock in a dusty medium, each species returns to the gas velocity over its own drag length, so a steady shock develops a layered relaxation structure that tests every coupling strength at once.
The stationary equations reduce to ordinary differential equations in $x$, and their numerical integration provides a semi-analytic reference \citep{BenitezLlambay2019}.

We drive an isothermal Mach~2 flow, with $v = 2$ and $c_{\rm s} = 1$, through a standing shock while carrying three dust species, each at the gas density.
The drag uses constant coefficients $K = 1$, 3, and 5, corresponding to stopping times $t_{{\rm s},k} = \rho_{k}/K_{k}$.
The dust triples the inertia of the mixture, so the effective sound speed of the fully coupled fluid halves.
The far-downstream compression therefore reaches the square of the effective Mach number, a factor of 16.
We initialize the two equilibrium states as a discontinuity on a domain of length 40 with 400 cells and relax the flow to a stationary profile by $t = 500$.

Figure~\ref{fig:dustyshock} overlays the numerical densities on the integrated reference.
The gas jumps at the front and then compresses smoothly as the dust catches up.
Each species relaxes over a length set by its drag coefficient, and all four profiles follow the reference through the entire layer.

\subsubsection{Drag Relaxation Under External Forcing} \label{sec:dustdamp}

\begin{figure*}[t]
\centering
\includegraphics[width=\textwidth]{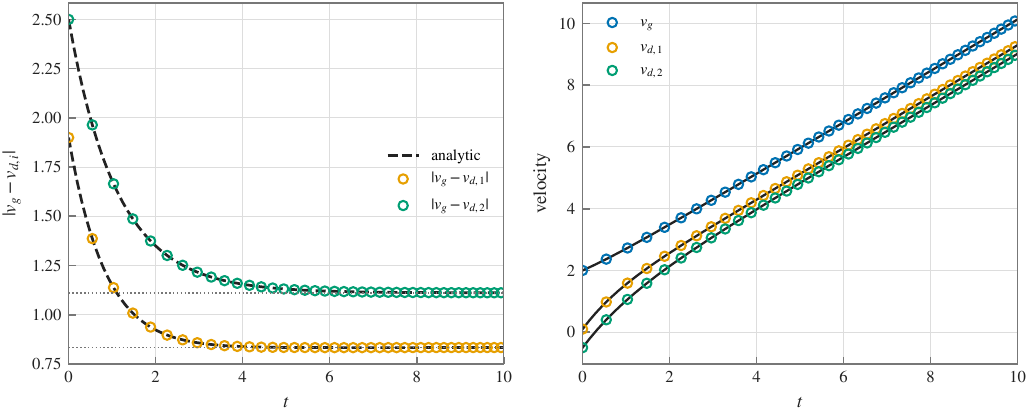}
\caption{Drag relaxation under a constant acceleration applied to the gas alone.
Left: gas-dust velocity differences; dashed lines show the exact solution, dotted lines the terminal drifts.
Right: the three velocities against time.}
\label{fig:dustdamp}
\end{figure*}

Section~\ref{sec:time} integrated drag inside the stages rather than operator splitting it, because splitting fails to reach the correct drift equilibrium when external forces act \citep{Krapp2024}.
Our final test checks that equilibrium directly.
A spatially uniform mixture carries gas of density 1 and two dust species of density 0.1 each, with stopping times 1 and $4/3$, starting from velocities 2, 0.1, and $-0.5$.
A constant acceleration $g = 1$ pushes the gas alone, so the mixture cannot settle into a comoving state.
The exact solution of the linear system instead approaches rigid acceleration at $\bar{a} = g\rho_{\rm g}/\rho_{\rm tot} \approx 0.83$, with each species trailing the gas by the terminal drift $v_{\rm g} - v_{{\rm d},k} = \bar{a}\,t_{{\rm s},k}$.

Figure~\ref{fig:dustdamp} shows both views of the relaxation.
The velocity differences decay onto the predicted terminal drifts, and the three velocities settle onto parallel linear growth at $\bar{a}$, matching the matrix-exponential solution throughout.
The automated suite repeats this problem in the stiff regime, with the timestep exceeding the shortest stopping time by an order of magnitude, and recovers the same drift equilibrium to better than one part in $10^{8}$.
This result confirms the L-stability of the implicit stages in a regime the resolved runs shown here cannot reach.
\rev{The suite also measures the temporal order of the drag update on an unforced variant of this problem, in which the mixture relaxes to its center-of-mass velocity.
Halving the step twice reduces the momentum error by factors of 3.5 and 3.7, close to the factor of 4 expected at second order.}

\section{Performance} \label{sec:performance}

Section~\ref{sec:intro} asked whether a dynamic language can deliver the performance and scalability expected of a production astrophysics code on modern GPU hardware.
In this section, we test whether that goal has been achieved, presenting the results of single-device performance and weak scaling tests on data-center, consumer, and exascale hardware.

Every benchmark below runs in double precision.
Because Julia compiles specialized kernels on first execution, each measurement begins after a warm-up phase that absorbs this one-time cost and covers the following 2000 cycles.

\subsection{Single-Device Performance} \label{sec:perf_single}

\begin{table}[t]
\centering
\caption{Single-device throughput on the three-dimensional linear-wave benchmark, in Mzcps. \rev{CPU rows use all cores of the chip: one shared $128^{3}$ block on the desktop processors, and one MPI rank per $64^{3}$ block on the EPYC.}
\label{tab:single_device}}
\begin{tabular}{@{}lcc@{}}
\toprule
Device & Hydro & MHD \\
\midrule
NVIDIA A100 (40 GB) & 709 & 300 \\
AMD MI250X (one GCD) & 478 & 250 \\
NVIDIA RTX 4090 (24 GB) & 215 & 122 \\
\midrule
\rev{Apple M4 Max (14 cores)} & \rev{33.5} & \rev{14.7} \\
\rev{AMD Ryzen 9 5900X (12 cores)} & \rev{12.8} & \rev{6.8} \\
\rev{AMD EPYC 7763 (64 cores, MPI)} & \rev{64.5} & \rev{31.8} \\
\bottomrule
\end{tabular}
\end{table}

To test whether kernels compiled from one Julia source remain competitive on hardware from different vendors, we run the same benchmark, unchanged, on \rev{three GPUs}.
The set comprises an NVIDIA A100 (40~GB) in the Perlmutter system at NERSC; one Graphics Compute Die (GCD) of an AMD MI250X in Frontier, a test system at the Oak Ridge Leadership Computing Facility (OLCF) with the same nodes as the exascale system Frontier; and a consumer NVIDIA RTX 4090 (24~GB).
The software stack exposes each GCD of an MI250X as an independent device, so per-GCD rates are the natural unit on this hardware.

For a direct comparison between the static and dynamic language, we adopt the single-device benchmark of AthenaK \citep{Stone2026}.
The benchmark advances a three-dimensional linear wave across a periodic uniform grid; the smooth solution keeps the work per cell uniform, so the timing reflects the sustained cost of the full update loop.
Both the hydrodynamic and the MHD solver advance the wave using PLM reconstruction, the HLLE Riemann solver, and the \code{RK2} integrator.
On the A100 and the MI250X the grid holds $512\times512\times256$ cells, tiled into $128^{3}$ blocks; the smaller memory of the RTX 4090 restricts its grid to $256^{3}$ cells at the same block size.

Table~\ref{tab:single_device} lists the measured throughput in millions of zone-cycles per device-second (Mzcps); the metric counts each cell once per cycle regardless of the stage count of the integrator.
The A100 sustains 709~Mzcps in hydrodynamics and 300~Mzcps in MHD, and a single MI250X GCD reaches 478 and 250~Mzcps.
Across the three \rev{GPUs} the MHD update costs 1.8 to 2.4 times the hydrodynamic update.
The ratio reflects the additional induction equation, the constrained-transport update, and the more expensive Riemann problem.

These rates place \aether\ in the performance class of contemporary compiled-language GPU codes.
Running the same combination of PLM, HLLE, and \code{RK2} on the same problem in double precision, AthenaK reports 614 and 298~Mzcps for hydrodynamics and MHD on an A100, and 405 and 190~Mzcps on the closely related AMD MI250.
The rates in Table~\ref{tab:single_device} match or exceed these values on both architectures.
This comparison suggests that dynamic Julia code runs comparably to statically compiled C++ code, and possibly slightly faster.

We include the consumer RTX 4090 to test performance on hardware outside the data center.
Consumer GPUs execute double-precision arithmetic at $1/64$ of their single-precision rate, so a large penalty might be expected.
However, the measured rates of 215 and 122~Mzcps instead fall only a factor of 2.5 to 3.3 below the A100. This suggests that it is the memory bandwidth, not arithmetic, limits most kernels of a finite-volume update \citep{Grete2021}, and the bandwidth gap between the two cards is far smaller than their arithmetic gap.
A personal computer therefore supports serious development and moderate production runs, which extends the single-language workflow to hardware a student may already own.

\rev{In addition to the GPUs, we benchmark \aether\ on three CPUs, where the same kernels run through the CPU backend of \code{KernelAbstractions.jl}.}
\rev{The set comprises a 14-core Apple M4 Max in a Mac Studio, a 12-core AMD Ryzen 9 5900X desktop processor, and a 64-core AMD EPYC 7763 server processor.}
\rev{On the two desktop processors, all cores share a single $128^{3}$ block through the auto parallelism from \code{KernelAbstractions.jl}.}
\rev{On the EPYC, a $256^{3}$ grid is divided into $64^{3}$ blocks, with one MPI rank per core and block.}
\rev{As for performance, the M4 Max sustains 33.5 and 14.7~Mzcps for hydrodynamics and MHD, and the 5900X sustains 12.8 and 6.8~Mzcps.}
\rev{The EPYC reaches 64.5 and 31.8~Mzcps.}
\rev{On every CPU the MHD update costs 1.9 to 2.3 times the hydrodynamic update, within the range measured on the GPUs.}

\subsection{Weak Scaling} \label{sec:perf_scaling}

\begin{figure}[t]
\centering
\includegraphics[width=\columnwidth]{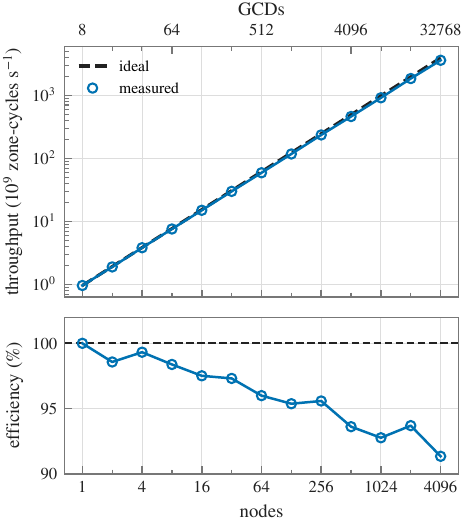}
\caption{Weak scaling of the driven-turbulence benchmark on Frontier.
Top: aggregate throughput against node count, with the dashed line marking ideal scaling from the single-node rate; the upper axis counts GCDs.
Bottom: parallel efficiency relative to one node.}
\label{fig:weak_scaling}
\end{figure}

To test parallel scaling at the scale of production simulations, we perform a weak scaling test on the Frontier system at OLCF.
Each Frontier node carries four MI250X accelerators, hence eight GCDs, and one MPI rank drives each GCD.
The workload per device stays fixed as the machine grows, so any increase in the time per cycle measures communication overhead and load imbalance.
The benchmark evolves driven isothermal MHD turbulence under stochastic forcing, so the timed loop includes the forcing kernels, the timestep reduction, and every communication step.
\rev{Each GCD carries a fixed load of $1.68\times10^{7}$ cells, and the node count grows in powers of two from 1 to 4096, so the largest run updates $5.5\times10^{11}$ cells on 32,768 GCDs, or 16,384 MI250X accelerators.}

Figure~\ref{fig:weak_scaling} shows the aggregate throughput and the parallel efficiency, the single-node time per cycle divided by its value at each node count.
\rev{Throughput rises from $9.6\times10^{8}$ zone-cycles per second on one node to $3.6\times10^{12}$ on 4096 nodes.}
\rev{Efficiency declines gradually with node count and remains above 92\% through 2048 nodes.}
\rev{The 4096-node run, the largest in the series, retains 91.3\%.}
\rev{To our knowledge, this is the largest GPU run of a Julia code to date.}
\rev{The previous largest, a stencil benchmark on Frontier, ran on 4,096 GCDs and met MPI-layer failures when scaled to 32,768 \citep{Godoy2023workflow}.}
\rev{Other documented Julia GPU runs reach 2,197 GPUs on Piz Daint \citep{Omlin2024} and 768 GPUs on Perlmutter \citep{Silvestri2023}.}
A solver written in Julia therefore sustains competitive per-device throughput and near-ideal weak scaling in large-scale runs.

\begin{figure*}[t]
\centering
\includegraphics[width=\textwidth]{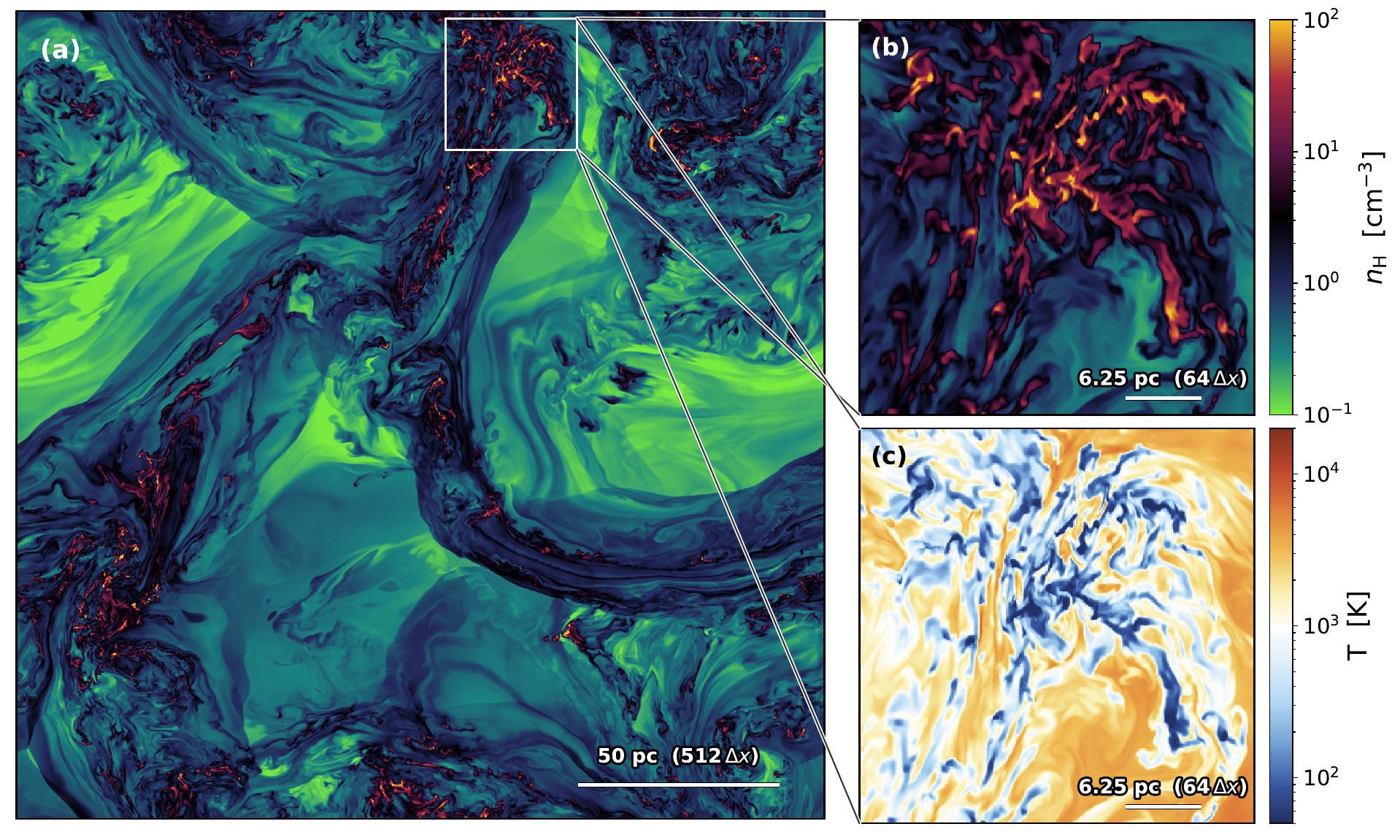}
\caption{Slice of the hydrogen number density in the saturated state of the multiphase turbulence run (a).
Zoomed-in views of the boxed region show the cold dense filaments and the surrounding warm gas in number density (b) and temperature (c).
Scale bars give lengths in parsecs and in grid cells.}
\label{fig:ism}
\end{figure*}

\section{Example Applications} \label{sec:applications}

The previous sections validate individual algorithms and measure performance on controlled workloads.
Here we show how the components of \aether\ work together in full production simulations.
The first is a $2048^{3}$ simulation of driven multiphase turbulence in the interstellar medium; the second follows the vertical shear instability in a protoplanetary disk containing 48 dust species.

\subsection{Multiphase Interstellar Turbulence} \label{sec:app_ism}

Compressible turbulence has long driven the largest uniform-grid calculations in computational fluid dynamics.
Studies of the universality of the cascade reached $10{,}048^{3}$, resolving the sonic scale \citep{Federrath2021}, and the magnetized counterpart of that measurement runs at $10{,}080^{3}$ \citep{Beattie2025}.

Behind these isothermal idealizations, the simulated gas is usually interpreted as the cold, dense component of the turbulent interstellar medium, where rapid cooling keeps the gas isothermal.
A more comprehensive model, however, must include its multiphase nature.
Heating and cooling maintain a warm phase near $10^{4}$~K and a cold phase near $10^{2}$~K in rough pressure balance \citep{Wolfire1995, Wolfire2003, HeilesTroland2005}, while turbulence continually drives gas across the thermally unstable branch between them \citep{KoyamaInutsuka2002}.
The rapid heating and cooling change the properties of the turbulence itself: the density and pressure statistics, and the energy spectra may depart from the isothermal case \citep{Kritsuk2017}.
Recent high resolution multiphase simulations have been used to study the plasmoid instability \citep{Fielding2023}, the polarized Galactic dust foreground \citep{2025PhRvD.112j1302H, Kritsuk2026}, and phase-dependent magnetic coherence \citep{Butsky2026}.

In such simulations, the warm phase remains transonic while the cold phase, whose sound speed is ten times lower, reaches $\mathcal{M}_{\rm s} \gtrsim 10$ across shocks connecting densities below $1~{\rm cm^{-3}}$ to above $100~{\rm cm^{-3}}$.
There the internal energy is only a percent of the total (Equation~\ref{eq:total_energy}), reconstruction overshoots could drive the pressure negative, and the pressure floor overwrites the temperature. As a result, the numerical scheme no longer evolves the thermodynamics of the gas correctly.
The dual-energy formalism solves this problem: it keeps the pressure positive and accurate even when the internal energy is a small fraction of the total.

Using this formulation, we evolve a 200~pc periodic box on a $2048^{3}$-cell grid with PPM reconstruction, the LHLLD solver, and \code{RK2}, assuming $\gamma = 5/3$.
Implicit cooling and heating act with fully solenoidal driving at $1 \le kL/2\pi \le 2$ on an initially uniform medium, $n_{\rm H} = 1~{\rm cm^{-3}}$, threaded by a uniform field.
Driving power and field strength put the saturated warm phase near $\mathcal{M}_{\rm s} \approx 1.2$ and $\mathcal{M}_{\rm A} \approx 2$.

The simulation was run on 256 Frontier nodes for two warm-phase sound-crossing times and completed in about four hours of wall clock.

Figure~\ref{fig:ism} shows a slice of the resulting density and temperature fields.
The dual-energy closure holds both phases at their equilibrium temperatures, from the volume-filling warm gas down to cold knots a few cells across.
Their boundaries stay sharp at the grid scale.
Between them, gas crossing the unstable range has no equilibrium to return to, and the closure alone preserves its thermal history.
A full analysis of the phase structure and of the synthetic polarization statistics will be presented elsewhere.

\begin{figure*}[t]
\centering
\includegraphics[width=\textwidth]{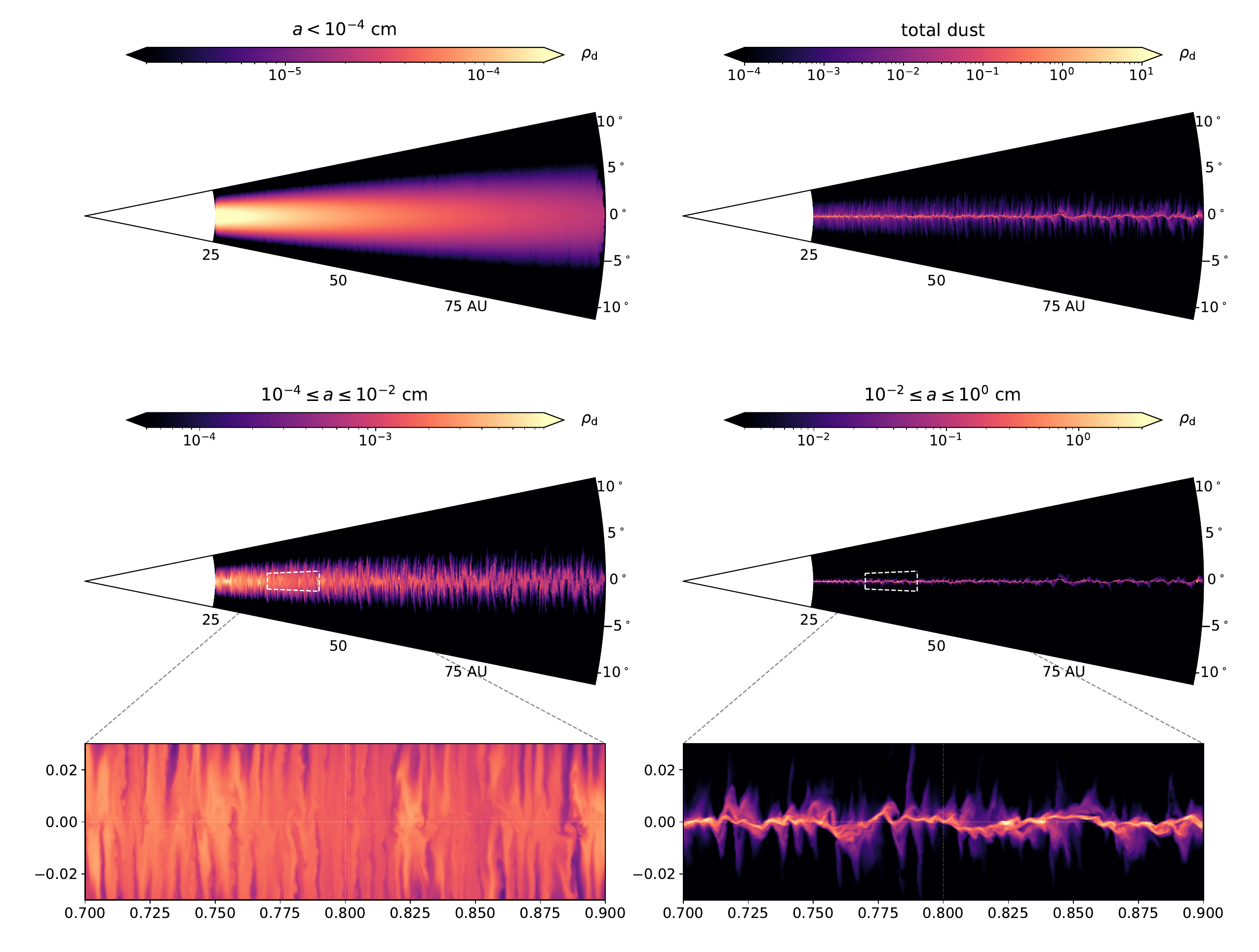}
\caption{Dust density in the 48-species VSI run after saturation of the instability.
Top: meridional maps of the dust density, in code units, summed over three grain-size bins ($a < 10^{-4}$~cm, $10^{-4} \le a \le 10^{-2}$~cm, $10^{-2} \le a \le 1$~cm) and over all species.
The wedges cover the full domain, 25 to 100~au in radius and $\pm 11.5^{\circ}$ about the midplane.
Bottom: magnifications of the boxed regions of the two larger-size bins, spanning $r \in [0.70, 0.90]\,r_{0}$ (35 to 45~au) and $|z| \le 0.03\,r_{0}$ ($\pm 1.5$~au).}
\label{fig:vsi}
\end{figure*}

\subsection{Vertical Shear Instability with 48 Dust Species} \label{sec:app_vsi}
In protoplanetary disks, the dynamics of dust depends not only on its total mass but on how that mass is distributed over grain size.
Grains of different sizes settle, drift, and are stirred at different rates.
The same distribution sets the opacity, and with it the thermal structure of the disk \citep{Testi2014, Birnstiel2016}.

Dust-driven instabilities respond to the full distribution rather than to a mean grain size: growth rates of the streaming instability change qualitatively once a size spectrum replaces a single species \citep{Krapp2019, Schaffer2018, ZhuYang2021, YangZhu2021, McNally2021, Rucska2023}.
The vertical shear instability \citep[VSI;][]{Nelson2013, StollKley2014} stirs grains at rates that depend on their size \citep{StollKley2016, Flock2017, Flock2020}, while the dust in turn alters the cooling rate on which the instability depends \citep{Dullemond2022, Pfeil2023, FukuharaOkuzumi2024}.

Grains also coagulate and fragment on timescales comparable to settling and drift.
Evaluating the collision kernel in every cell is expensive in CPU time, so such modeling has stayed largely one-dimensional \citep{Brauer2008, Birnstiel2010, Estrada2016, Stammler2022}, and couplings to multidimensional hydrodynamics remain rare \citep{Drazkowska2019, Laune2020, Li2020, Ho2024}.
Here we demonstrate the transport side of that coupling at production scale: the dust module of \aether\ evolves the full grain size distribution through a disk instability.
In the future, on-the-fly coagulation could then be added as a user-defined source term within the source framework of \aether\ \rev{\citep{Yang2026_inprep}}, compatible with current methods \citep{Birnstiel2010, Lombart2021, Lombart2024, Yang2026}.

We simulate the VSI with 48 dust species, in the axisymmetric equilibrium disk of \citet{Nelson2013} around a $GM = 1$ point mass: aspect ratio $h = 0.05$, midplane density $\propto R^{-2}$, and $c_{\rm s}^{2} = h^{2}\,GM/R$ on cylindrical radius $R$.
Gravity enters as a user-defined source term.
The VSI requires rapid cooling toward a locally isothermal state \citep{LinYoudin2015}.
A per-stage hook therefore resets the internal energy of the $\gamma = 7/5$ gas to this profile, with drag heating disabled.

The species sample grain radii log-uniformly from $10^{-6}$ to 1~cm, with mass weights following the \citet{Mathis1977} distribution $n(a) \propto a^{-3.5}$ at total dust-to-gas ratio $Z = 0.01$.
For $r_{0} = 50$~au around a solar mass, with the minimum-mass solar nebula surface density $\Sigma_{\rm g} = 4.8~{\rm g\,cm^{-2}}$ \citep{Hayashi1981} and material density $1.25~{\rm g\,cm^{-3}}$, Epstein drag gives midplane Stokes numbers from $4.1\times10^{-7}$ to $0.41$.

The axisymmetric spherical-polar mesh spans $r \in [0.5, 2]\,r_{0}$ and $\theta = \pi/2 \pm 4h$ with $3072 \times 1024$ uniform cells, 102 radial and 128 meridional per scale height.
This exceeds published convergence requirements for the instability \citep{FloresRivera2020, Manger2020}.
We combine PPM with HLLC for the gas, the pressureless HLL flux for the dust, and \code{IMEX2P} at a Courant factor of 0.4.

The dust starts from the drift equilibrium of \citet{Nakagawa1986}, generalized to many species \citep{BenitezLlambay2019}, with each species settled to a scale height $h_{\rm d} = h \sqrt{\delta_{0}/(\delta_{0} + {\rm St})}$ at $\delta_{0} = 10^{-3}$ \citep{Dubrulle1995, YL2007} and floored at three cells.
A random meridional velocity perturbation of $10^{-3}\,c_{\rm s}$ seeds the instability.
Dirichlet boundaries hold all four faces at the initial equilibrium, so the outer face resupplies inward-drifting dust and the disk reaches a drift-through steady state.
The run advances 125 orbits at $r_{0}$, about 44~kyr in $1.54\times10^{6}$ cycles, on two Frontier nodes in twelve hours of wall clock.
With 49 fluids per cell it sustains $1.18\times10^{8}$ zone-cycles per second, or $3.6\times10^{8}$ per-species cell updates per GCD-second, on par with the single-fluid rates of Table~\ref{tab:single_device}.

Figure~\ref{fig:vsi} shows the resulting dust density in three size bins and in total.
The smallest grains (${\rm St} \lesssim 4\times10^{-5}$) spread through the full domain and trace the gas.
Intermediate sizes follow the corrugated columns of the VSI body modes, resolved across dozens of cells in the left zoom.
The largest grains settle into a thin midplane layer that the corrugation bends and locally disrupts (right zoom).
Edge-on disks show the same stratification of well-mixed small grains around a thin pebble layer.
A quantitative analysis of the size-dependent scale heights and drift-through fluxes will follow in a dedicated paper.

\section{Discussion and Summary} \label{sec:summary}

In this paper, we presented \aether\ and a structured workflow for developing a production simulation code with coding agents.
\rev{In the workflow of Section~\ref{sec:workflow}, the human divided development into individual agent sessions and curated the shared design documents.
Repository memory connected these sessions, while scientific decisions and verification remained under human oversight.}

\aether\ itself implements finite-volume compressible MHD with multifluid dust in Julia.
The code is verified against standard test problems, matches the throughput of contemporary C++ GPU codes, and holds \rev{above 91\% parallel efficiency on 32,768 GCDs} of Frontier. 
Two production applications showcase ran at $2048^{3}$ cells and a 2D run with 48 dust species.

\aether\ currently targets applications in astrophysical turbulence and its coupling to multiphysics, from the multiphase interstellar medium to protoplanetary disks.
Beyond these applications, Julia's single-language design gives the solver access to a wider scientific ecosystem.
The stiff integrators of \code{DifferentialEquations.jl} and their GPU ensemble variants \citep{RackauckasNie2017, Utkarsh2024} could evolve astrochemical networks in every cell \citep[e.g.,][]{KROME}.
Large networks may make direct integration too expensive.
A neural surrogate built with \code{Lux.jl} \citep{Pal2023Lux} could approximate the local solve and be evaluated on CPU or GPU arrays within the same Julia program.
Fluid solvers increasingly incorporate such models \citep{Kochkov2021, Li2021, Lu2021}, and stiff astrochemistry has already been emulated with autoencoders and neural operators \citep{Grassi2022, BrancaPallottini2024}.
On-the-fly grain coagulation poses a similar cell-local problem and could use either strategy.

These extensions suggest a broader model for simulation software.
A Julia simulation can be assembled from interoperable scientific libraries instead of growing as a closed executable.
The fluid solver, local physics modules, analysis routines, and learned components can share data structures within one program.
Researchers can develop a new physical model interactively, couple it directly to the time-stepping loop, deploy it on CPUs or GPUs, and analyze the results without rewriting the model in another language.
This structure brings prototyping, production simulation, and analysis into a unified Julia workflow.
The results presented here show that such integration can retain the performance and scalability required for production astrophysics.

\rev{A parallel shift is under way in how such codes are built.
The record of Section~\ref{sec:record} suggests a transition from human-coordinated sessions built around ReAct-style loops, in which one agent alternates reasoning and tool calls \citep{Yao2023}, to workflows managed by a coordinating agent.
Claude Code's dynamic workflows \citep{Anthropic2026} and DeepSeek's Super-agent prototype \citep{DeepSeek2026} illustrate this direction.
In this pattern, a coordinating agent maintains the dialogue with the human, divides objectives into tasks, delegates them to workers, and integrates their results.
Shared repository memory could connect these workers and preserve decisions across successive workflows.
The coordinator could also maintain that memory by reconciling findings, updating design documents, and promoting lessons into reusable rules.
Extending this coordination to simulation execution and diagnosis could shift human effort toward scientific questions, compute budgets, and falsification criteria.
Whether such a system can keep a scientific code coherent, verified, and extensible across years of development by many contributors remains an open question as \aether\ evolves.}

\section*{Acknowledgments}
The author thanks Alexei G. Kritsuk for insightful discussions and constructive comments on the manuscript. The author also thanks the organizers of the 2026 LANL Julia for Science Training Workshop for the valuable technical insights gained during the workshop. This research used resources of the National Energy Research Scientific Computing Center, a DOE Office of Science User Facility supported by the Office of Science of the U.S. Department of Energy under Contract No. DE-AC02-05CH11231, using NERSC awards ASCR-ERCAP0036667 and FES-ERCAP0036102. Additional computational and storage resources were provided through the Director's Discretion allocation at Oak Ridge National Laboratory (ORNL), award AST193. This work was supported in part by grant NSF PHY-2309135 and Simons Foundation International SFI-MPS-T-Institutes-00021142, LB. It has also benefited from the Mitchel Postdoctoral Scholar Career Development Fund. \rev{We thank Jia-Xin Zhang for providing the performance number of Apple M4 Max Chip. We thank the anonymous referee for a constructive review, which helped to improve the overall quality and clarity of the paper.}

\section*{\rev{Software and Data Availability}}
\rev{The version of \aether\ used in this work is archived on Zenodo at \dataset[doi:10.5281/zenodo.24047556]{https://doi.org/10.5281/zenodo.24047556}.
The Jupyter notebooks that reproduce most of the validation figures are archived at \dataset[doi:10.5281/zenodo.24047599]{https://doi.org/10.5281/zenodo.24047599}.}

\software{\aether\ \rev{\citep{Ho2026Aether}}, \rev{Aether-Example \citep{Ho2026AetherExamples}}, Julia \citep{Bezanson2017}, KernelAbstractions.jl \citep{Churavy2021}, CUDA.jl \citep{Besard2019}, AMDGPU.jl \citep{Samaroo2026}, MPI.jl \citep{Byrne2021}, ADIOS2 \citep{Godoy2020}, ADIOS2.jl, Matplotlib}

\section*{\rev{AI Disclosure Statement}}
\rev{The author used Claude Cowork and Claude Code (Anthropic; Fable 5.0 and 5.1, Opus 4.8 and 5.5) to develop the agent framework and implement \aether\ (Section~\ref{sec:workflow}).
Codex (OpenAI; ChatGPT 5.6 Sol and 6.0 Astra) was also used for both tasks.
During manuscript preparation, the author used Claude Code (Fable 5.0) and Antigravity (Google; Argon 4) to assist with analysis scripting, numerical verification, and prose editing.
All scientific content, methodology, and interpretations are the author's own.
The author reviewed and verified all code, outputs, and text, and takes full responsibility for the integrity and accuracy of the work.}

\bibliographystyle{aasjournal}
\bibliography{all}

\end{document}